\documentclass[iop,apj]{emulateapj}
\slugcomment{{\sc Accepted to ApJ:} March 16, 2015}

\usepackage{amsmath}
\usepackage{wasysym}
\usepackage{natbib}
\usepackage{url}

\begin{document}

\title{Mapping the Outer Edge of the Young Stellar Cluster in the Galactic Center}
\author{M. St\o stad\altaffilmark{1}, T. Do\altaffilmark{2,3}, N. Murray\altaffilmark{4}, J. R. Lu\altaffilmark{5}, S. Yelda\altaffilmark{3}, A. Ghez\altaffilmark{3}
}

\altaffiltext{1}{Department of Astronomy \& Astrophysics, University of Toronto, 50 St. George Street, Toronto M5S 3H4, ON, Canada; morten.stostad@mail.utoronto.ca}
\altaffiltext{2}{Dunlap Institute for Astronomy and Astrophysics, University of Toronto, 50 St. George Street, Toronto M5S 3H4, ON, Canada}
\altaffiltext{3}{UCLA Department of Physics and Astronomy, Los Angeles, CA 90095-1547, USA}
\altaffiltext{4}{Canadian Institute for Theoretical Astrophysics, University of Toronto, 60 St. George Street, Toronto, M5S 3H8, ON, Canada.}
\altaffiltext{5}{Institute for Astronomy, University of Hawaii, Honolulu, HI96822, USA}


\begin{abstract}

We present new near-infrared spectroscopic observations of the outer edges of the young stellar cluster around the supermassive black hole at the Galactic center. The observations show a break in the surface-density profile of young stars at $\sim 13\arcsec$ (0.52 pc). These observations spectroscopically confirm previous suggestions of a break based on photometry. Using Gemini North's Near-Infrared Integral Field Spectrometer (NIFS) we are able to detect and separate early- and late-type stars with a $75\%$ completeness at $\emph{K}_s = 15.5$. We sample a region with radii between $7\arcsec$ to $23\arcsec$ (0.28 pc to 0.92 pc) from Sgr A*, and present new spectral classifications of 144 stars brighter than $\emph{K}_s = 15.5$, where 140 stars are late-type ($> 1$ Gyr) and only four stars are early-type (young, 4-6 Myr). A broken power-law fit of the early-type surface-density matches well with our data and previously published values. The projected surface-density of late-type stars is also measured and found to be consistent with previous results. We find that the observed early-type surface-density profile is inconsistent with the theory of the young stars originating from a tightly bound infalling cluster, as no significant trail of young stars is found at radii above $13\arcsec$. We also note that either a simple disk instability criterion or a cloud-cloud collision could explain the location of the outer edge, though we lack information to make conclusive remarks on either alternative. If this break in surface-density represents an edge to the young stellar cluster it would set an important scale for the most recent episode of star formation at the Galactic center.

\end{abstract}



\section{Introduction}

Sagittarius A* (Sgr A*), the supermassive black hole at the center of our galaxy, is surrounded by one of the richest concentrations of massive stars in the Milky Way. Star formation in a region so close to such a massive object is theoretically difficult due to the extreme tidal forces from Sgr A*, so the detection of young stars ($<6$Myr) in the inner half parsec of the Galactic center (GC) in the early 1990s was remarkable \citep{allen1990,krabbe1991,krabbe1995,blum1995}. Since then nearly 200 young O- and B-stars have been confirmed in the inner parsec, almost 5\% of the number of late-type giants in the region \citep{najarro1997,ghez2003,paumard2006,lu2009,bartko2010,genzel2010,do2013,yelda2014}. 

The inner regions of this young stellar cluster have been extensively mapped with the use of integral field spectroscopy and high-resolution imaging. These methods yield positions, radial velocities, and proper motions that are used to model the orbital dynamics of the young stars \citep{paumard2006,do2009,bartko2010,do2013}. The cluster extends azimuthally to at least $12\arcsec$ (0.48 pc) from Sgr A*. No firm outer edge has been detected, largely due to lack of spectroscopic data with high spatial resolution outside $12\arcsec$. Additionally, the existence of a kinematically distinct clockwise disk consisting of young stars from the cluster has been confirmed by several sources \citep{levin2003,paumard2006,lu2009,bartko2009,lu2013,yelda2014}. The fraction of young stars that are also members of this disk has been estimated to be between 20-55\% \citep{bartko2009,yelda2014}. The disk has an inner edge at $0.\arcsec8$ from Sgr A* and could extend the full radius of the young stellar cluster, but again, the outer edge is largely unexplored. \citet{bartko2010} presents spectroscopic data out to $25\arcsec$, but with low azimuthal coverage, relatively low completeness and a primary focus on the stars inside $12\arcsec$. \citet{buchholz2009} and \citet{nishiyama2013} present maps of candidate early-type stars up to 2.5pc ($\sim 60\arcsec$) -- however, their work is done without the use of spectroscopic analysis, which has the disadvantage of a spurious detection rate of roughly 20\% and a smaller magnitude range than in this work. As such, only a limited amount of spectroscopic analysis has been done outside the central 0.5pc of the Galactic center, and the physical scale of recent star formation in the Galactic center is not well measured.

The origin of the young stars at the Galactic center is an active area of research. Star formation via gravitational collapse in regions within 3 parsec ($75\arcsec$) of Sgr A* would require gas about five orders of magnitude denser than what is currently observed in the Galactic center \citep[roughly $10^3$ to $10^8$ cm$^{-3}$,][]{jackson1993,christopher2005,montero2009}. Stars forming outside the stellar cluster and individually migrating inwards is also unlikely, as the two-body relaxation times from such distances are longer than the star lifetimes. Currently the two most prominent theories for star formation in the Galactic center are \emph{in situ} formation and the infalling cluster hypothesis. 

\emph{In situ} star formation scenarios focus on a massive self-gravitating gaseous disk around Sgr A* reaching the required densities to overcome tidal shear and form stars. It is currently the favored theory of GC star formation largely because the expected density profile matches well with measured values. The theoretical surface-density profile for \emph{in situ} formation falls off roughly as $r^{-3/2}$ \citep{lin1987,paumard2006}, and the observed density profile falls off as $r^{-1.7}$ \citep{paumard2006,lu2009,bartko2010,yelda2014}. Other disk features, for example the disk thickness and the well-defined inner edge, also strengthen the \emph{in situ} theory \citep{paumard2006,bartko2010,lu2013}. One of the major remaining issues with the \emph{in situ} model is the need of a better understanding of the origin of the gas disk itself. One theory is that two giant molecular clouds may have collided, lost their angular momentum, and formed a star-forming gaseous disk around Sgr A* \citep{hobbs2009}. Cloud-cloud collisions may also be able to explain the large off-disk stellar population, which is difficult to reconcile with many other star formation scenarios.

The infalling cluster hypothesis theorizes that a tightly gravitationally bound cluster formed young stars at large radii where the tidal influence of Sgr A* is weak. This cluster then quickly migrated inwards due to dynamical friction, thereby explaining the young stars in the inner parsec \citep{gerhard2001,mcmillan2003,kim2003}. Such a scenario would leave a trail of young stars further out with a shallow surface-density profile, falling as $r^{-0.75}$ \citep{berukoff2006}. This is much shallower than the current observed density profile. However, we can presently only observe the most massive early-type stars in the GC, which has led to some concerns that mass segregation might make the density profile appear steeper than it is. As a result even a steep density profile (like the one currently observed) is insufficient evidence in ruling out the infalling cluster hypothesis. A stronger result would be in form of an edge to the young stellar cluster -- such an edge would be a definite signature of the \emph{in situ} theory, as it indicates either the outer boundary of the original gas disk or the radius at which the disk was no longer dense enough to form stars. Until now, however, it has been difficult to completely rule out the possibility of an infalling cluster due to the low number of spectroscopic surveys in regions further away from the GC than $12\arcsec$ (0.5 pc).

We have carried out a new high-resolution spectroscopic analysis of a region largely outside the inner $12\arcsec$ (0.48pc), spanning from $7\arcsec-23\arcsec$ ($0.28-0.92$pc), to measure surface-density profiles of both the young and the old stars. Most of the stars in the observed region have not previously been spectroscopically analysed, and only \citet{bartko2010} has high-resolution spectroscopic observations extending further out (projected distances of up to $25\arcsec$, though with a lower completeness and low azimuthal coverage). Our data sets, observation techniques, data reduction and spectroscopic classification are detailed in Section~\ref{sec:obs}. The completeness of our observations and number counts are shown in Section~\ref{sec:comp}. The remaining data analysis and our results are presented in Section~\ref{sec:res}, and finally the discussion and conclusion are in Section~\ref{sec:dis} and~\ref{sec:conc}.


\section{Observations and Data Reduction}
\label{sec:obs}

\subsection{Observations}

\begin{figure*}
\centering
\includegraphics[width=6in]{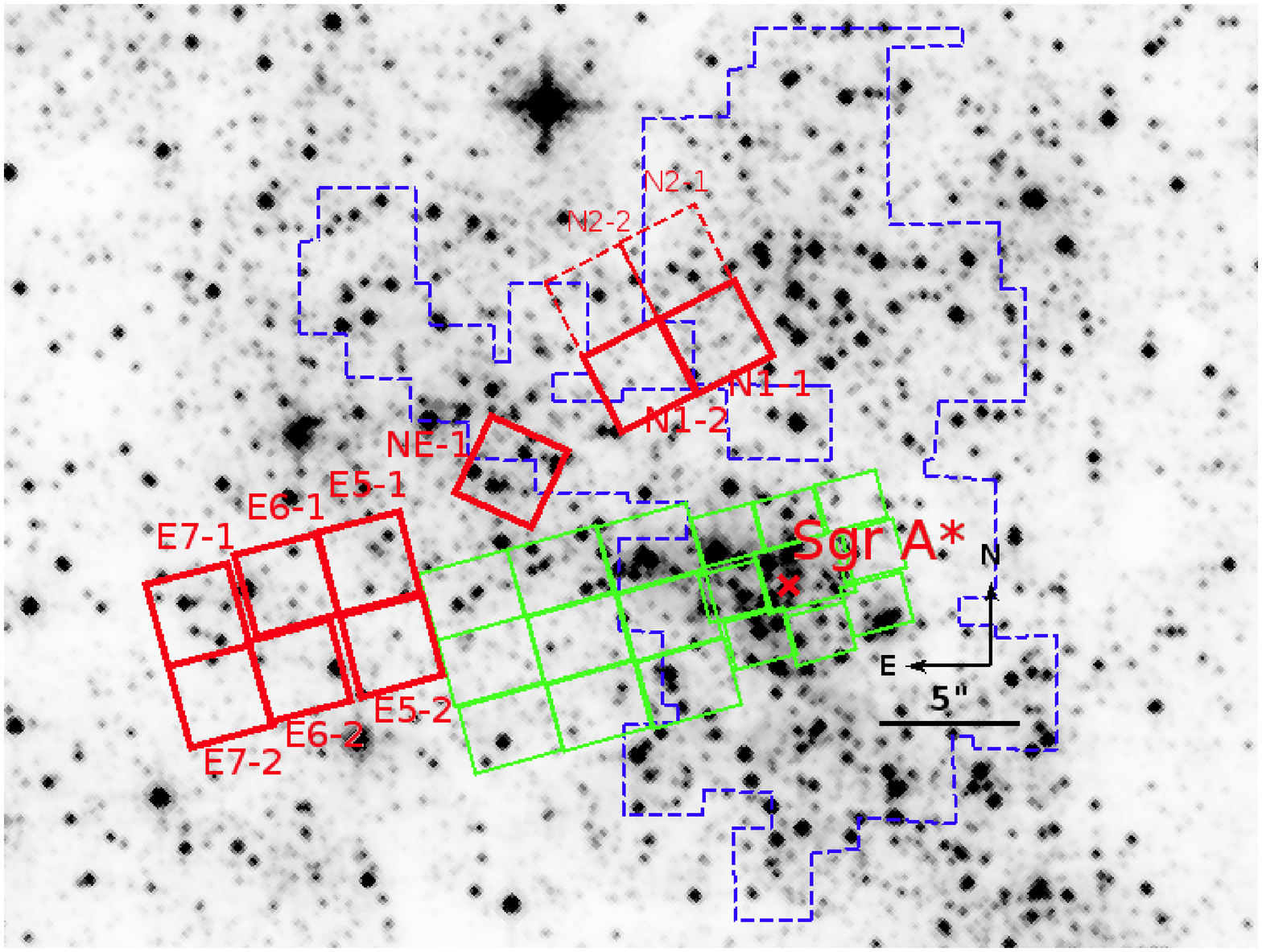}
\caption{The red regions are the new Gemini NIFS regions observed in this paper, while the green regions closer to Sgr A* is the data from \citet{do2013}. The blue regions are ones observed by \citet{bartko2010}, with lower completeness than the observations from this paper. The photometric \citet{buchholz2009} observations used a $40\arcsec$x$40\arcsec$ region centered roughly on Sgr A*. The background image is from HST observations of the nuclear star cluster (GO-12182, PI Do).}
\label{fig:starlocs}
\end{figure*}

Observations were obtained with Gemini-North's Near-Infrared Integral Field Spectrometer (NIFS) between May 2012 and May 2014 with natural guide-star and laser-guide star adaptive optics.\footnote{Gemini observations GN-2012A-Q-41 and GN-2014A-Q-71.} NIFS provides a 2040 pixel wavelength spectrum in the K-band (1.965-2.430 $\mu$m) for a continuous $3\arcsec$ x $3\arcsec$ spatial field (62 x 60 pixels). The spectral resolution is $\sim5000$ with a typical spatial resolution of 115-165 mas. The radial extent of the survey is from $7.5\arcsec$ to $23\arcsec$ in projected distance from the Galactic center. The fields are separated into two regions, one north and one east of Sgr A*, for an approximate total surface area coverage of 99 sq. arcsec.\footnote{Two nights were discarded before science analysis due to poor AO correction (regions N2\_1 and N2\_2). The total area coverage of the analyzed data is 81 arcsec$^2$.} Each field was observed five or six times with an exposure time of 600 s per frame and a small ($\sim0.\arcsec2$) dither pattern. The fields are situated so as to avoid areas already examined by previous deep spectroscopic surveys and to examine either the projected location of the clockwise (CW) young stellar disk (eastern fields) or the region perpendicular to the CW young stellar disk (northern fields). The exception is the field NE1-1, which is between the two regions -- for the purpose of this paper it will be considered as part of the northern regions. The locations of the fields, together with relevant observation regions from other papers, are indicated in Fig~\ref{fig:starlocs} -- information about each field is reported in Table~\ref{tab:obs}. 

We observe skies and two A-stars (HD155379, HIP96408) each night for calibration purposes. The Galactic center is densely populated with stars, and as such the science exposures could not be used for sky subtraction -- separate sky observations from the region RA 17:45:50.558, Dec -29:00:01.30 were used for this purpose, while the science exposures themselves were used for annulus subtraction around each star (see Section~\ref{sec:dat}). The A-stars were used to correct for atmospheric absorption lines. We also use two tip-tilt stars, one for the northern regions (USNO 0609-0602733, RA 17:45:42.287, Dec -29:00:36.80) and one for the southern regions (USNO 0690-0602749, RA 17:45:40.720, Dec -29:00:11.20).

\begin{deluxetable*}{lcclcccc}
\tablewidth{0pt}
\tabletypesize{\scriptsize}
\tablecolumns{8}
\tablecaption{Summary of NIFS Observations
\label{tab:obs}}
\tablehead{    
  \colhead{Name} &
  \colhead{Field Center\tablenotemark{a} ($\arcsec$)} &
  \colhead{Radius ($\arcsec$)} &
  \colhead{Date (UT)} &
  \colhead{FWHM (mas)\tablenotemark{b}} &
  \colhead{$N_{frames}$\tablenotemark{c}} &
  \colhead{Guide star} &
  \colhead{Avg extinction ($A_{\emph{K}_s}$)}
}
\startdata
E5-1	 & 15.34, 0.43	 & 15.34  & 2012-05-07	 & 135	 & 5	 & LGS & 2.58 \\ 
E5-2	 & 14.62, -2.45	 & 14.82  & 2013-05-23	 & 140	 & 5	 & LGS & 2.75 \\ 
E6-1	 & 18.24, -0.45	 & 18.25  & 2013-05-30	 & 150	 & 5	 & LGS & 2.67 \\ 
E6-2	 & 17.65, -3.35	 & 17.97  & 2012-05-05	 & 115	 & 5	 & LGS & 2.73 \\ 
E7-1	 & 21.27, -1.35	 & 21.31  & 2012-05-12	 & 150	 & 5	 & NGS & 2.68 \\ 
E7-2	 & 20.57, -4.35	 & 21.02  & 2012-05-12	 & 145	 & 6	 & NGS & 2.66 \\ 
N1-1	 & 2.73, 8.67	 & 9.09   & 2012-05-11	 & 165	 & 5	 & NGS & 2.61 \\ 
N1-2	 & 5.36, 7.24	 & 9.01   & 2012-05-11	 & 165	 & 5	 & NGS & 2.75 \\ 
NE1-1	 & 10.08, 3.80	 & 10.77  & 2014-05-14	 & 135	 & 5	 & LGS & 2.41 \\ 
N2-1\tablenotemark{d}	 & 4.23, 11.35	 & 12.11 & 2012-05-13	 & 225	 & 5	 & NGS & 2.63 \\ 
N2-2\tablenotemark{d}	 & 6.84, 9.88	 & 12.02 & 2012-05-13	 & 350	 & 5	 & NGS & 2.76 
\enddata
\tablenotetext{a}{R.A. and decl. offset in projected distance from Sgr A* (R.A. offset is positive to the east).}
\tablenotetext{b}{Average FWHM found from two-dimensional Gaussian fits to the PSF used in \emph{Starfinder} (PSF extracted from relatively isolated stars in the field).}
\tablenotetext{c}{Each frame has an integration time of 600 s.}
\tablenotetext{d}{Discarded prior to analysis due to poor AO correction.}
\end{deluxetable*}


\subsection{Data Reduction}
\label{sec:dat}

Data reduction of the raw science exposures was done using a modified version of the NIFS pipeline supplied by Gemini.\footnote{The original pipeline can be found at \url{http://www.gemini.edu/sciops/instruments/nifs/?q=node/10356}. Our modified pipeline will be available in the ApJ online supplementary material.} The pipeline performs sky subtraction, dark subtraction, flat field correction, and corrects for optical distortion on the detector. It also uses the spectra of the two A0V-stars to correct for atmospheric absorption. The A-star spectra are featureless except for a strong Br $\gamma$ absorption feature. To remove this feature we divide the observed spectrum by a theoretical Vega spectrum from the Kurucz models\footnote{\url{http://kurucz.harvard.edu/stars/VEGA/}} (that has been convolved to match the A-star spectral resolution). After the pipeline, the resulting science frames were combined for each field, aligning individual data cubes by using the positions of bright sources in each frame.

\begin{figure}
\centering
  \includegraphics[width=0.8\linewidth]{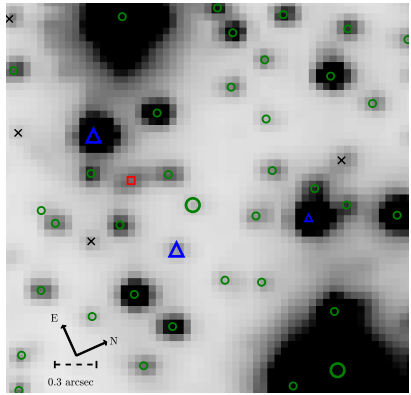}
  \caption{Data cube from region NE1-1, at roughly $11\arcsec$ (0.44 pc) distance from Sgr A*. All detected stars are marked, the different markers representing early-type stars (blue triangles), late-type stars (green circles), stars that could not be spectral typed (black crosses), and one foreground star (red square). The stars corresponding to the spectra shown in Figure~\ref{fig:spectra} have larger markers. Similar images of all the data cubes will be included in the ApJ online supplementary material.}
  \label{fig:image}
\end{figure}

To locate the stars the point-spread function (PSF) fitting program \textit{StarFinder} \citep{diolaiti2000} was used on collapsed versions of the science exposures. At times, \textit{Starfinder} would fail to detect very obvious stars even when a very low correlation threshold was used in the detection algorithm. Due to this we both used a very low correlation threshold ($CT=0.4$) and checked visually for obvious false detections or omissions afterwards. We find that \textit{Starfinder} detects simulated stars (at their correct magnitude) with a 50\% completeness down to $\emph{K}_s = 16.2$.  At fainter magnitudes false detections appear. We made an attempt to manually remove these, but detections fainter than $\emph{K}_s = 16.2$ should still be approached with some caution (especially those not cross-referenced to a detection from \citet{schodel2010}). More details about the detection completeness can be found in Section~\ref{sec:comp} and in Appendix~\ref{App:close}. Specifically it should be noted that these simulations lack the visual check for obvious omissions and false detections we make for the science exposures, and that this may lead to the simulations slightly underestimating photometric completeness (see Section~\ref{sec:photocomp}). Absolute coordinates were then assigned to the sources by comparing pixel positions to known stellar positions from HST observations of the nuclear star cluster (GO-12182, PI Do), which were aligned to 2MASS observations. The stellar coordinates are listed in Table~\ref{tab:early},~\ref{tab:late} and~\ref{tab:unknown} (Tables \ref{tab:late} and~\ref{tab:unknown} can be found in Appendix~\ref{App:tables}). One of the final data cubes, with all star positions marked, can be seen in Figure~\ref{fig:image}. The rest of these data cube images will be available in the ApJ online supplementary material.

After obtaining the stellar positions, their spectra were extracted by way of aperture photometry on each spectral channel. The flux from a circular aperture with a radius of 1.8 pixels ($\sim$90 mas) around the star was extracted. Furthermore, a median background spectrum with a 2 pixel annulus starting at 2 pixels from the center was subtracted from the spectrum of the star to remove the local background. Though this procedure may include very close stars in the background estimation, taking the median and manually checking stars for close companions helps to minimize this issue. Furthermore, for all close companions star-planting simulations were performed to check our performance in extracting the fainter spectrum. If a brighter star contaminated the spectrum (but not so much as to make the spectral type uncertain) it is noted in Table~\ref{tab:early} and \ref{tab:late}. (See also Appendix~\ref{App:close}).

The signal to noise ratio (SNR) per spectral channel was calculated with the region between 2.212 and 2.218 $\mu$m, which is relatively featureless in our spectra. We noted that spectrum quality of the brightest stars does not scale well with increasing flux at SNR\textgreater40 due to intrinsic lines in the late-type stellar spectrum. We were unable to locate any regions with less features in our high SNR late-type spectra, so most bright stars' SNR are only constrained to be above 40. We still expect SNR to scale with square root of star flux above this limit.

The $\emph{K}_s$-band magnitudes of the stars were found by matching detections to the photometry from \citet{schodel2010}.\footnote{\url{http://vizier.cfa.harvard.edu/viz-bin/VizieR?-source=J/A+A/511/A18}} Location and estimated magnitudes were used to search for matching stars, after which the $\emph{K}_s$ values from the \citet{schodel2010} measurements (if applicable) were used for further analysis. If no match could be found an approximate magnitude calculated from the spectra intensity was used. Additionally, we use an extinction map of the Galactic center (again from \citet{schodel2010}) to shift all magnitude measurements to a mean extinction value of $A_{\emph{K}_s} = 2.7$. The \citet{schodel2010} extinction map shows average $A_{K_s}$ values of between 2.41 and 2.75 between the different fields observed in this study (Table~\ref{tab:obs}). These values are about average for the extinction in this region. \citet{scoville2003} showed that the most eastern fields (near the circumnuclear disk) may have slightly higher extinction, though they use a different extinction law compared to \citet{schodel2010}. Regardless, the difference in extinction does not substantially affect our conclusions given the high completeness of this survey.


We extract a total of 375 spectra from the survey region. We utilize all 221 spectra from stars in the eastern fields for the analyses in this paper. In the northern fields, however, all data -- 26 spectra -- from the fields N2-1 and N2-2 are discarded due to very low completeness (see Section~\ref{sec:comp}). We are left with 128 usable spectra from the northern regions and 349 spectra in total.\footnote{This is before applying the magnitude cutoff -- see Section~\ref{sec:spec}.} The early-type stars are listed in Table~\ref{tab:early}, while the late-type stars are listed in Table~\ref{tab:late}. The stars that could not be spectral typed are listed in Table~\ref{tab:unknown}. Table~\ref{tab:late} and Table~\ref{tab:unknown} -- listing the late-type and unknown stars -- are in Appendix~\ref{App:tables}.

\begin{deluxetable*}{lccccccccccccrr}
\tabletypesize{\scriptsize}
\tablecolumns{15}
\tablecaption{NIFS Observations of Early-Type Stars
\label{tab:early}}
\tablehead{    
  \vspace{-0.12cm} &  &  &
  \colhead{$\Delta$R.A.\tablenotemark{a}} &
  \colhead{$\Delta$Decl.\tablenotemark{a}} &
  \colhead{R} &  &
  \colhead{CO 2.2935} &
  \colhead{Err} &
  \colhead{CO 2.3227} &
  \colhead{Err} &
  \colhead{Br$\gamma$ EW} &
  \colhead{Err\tablenotemark{b}} &  & \\
  \colhead{Name} \vspace{-0.15cm} &
  \colhead{Date} &
  \colhead{$\emph{K}_s$} &   &  &  &
  \colhead{SNR} &  &  &  &  &  &  &
  \colhead{$A_{\emph{K}_s}$} &
  \colhead{Contam} \\  &  &  & 
  \colhead{($\arcsec$)} &
  \colhead{($\arcsec$)} &
  \colhead{($\arcsec$)} &
  &
  \colhead{EW (\AA)} &
  \colhead{(\AA)} &
  \colhead{EW (\AA)} &
  \colhead{(\AA)} &
  \colhead{(\AA)} &
  \colhead{(\AA)} &  &
}
\startdata
N1-2-043	 & 2012-05-11	 & 12.8	 & 3.73	 & 7.15	 & 8.07	 & 67	 & 0.0	 & 0.3	 & 1.1	 & 0.2	 & 1.4	 & 0.4	 & 2.60	 & N \\ 
N1-1-017	 & 2012-05-11	 & 16.0	 & 4.08	 & 8.37	 & 9.31	 & 16	 & 1.4	 & 1.1	 & 1.4	 & 0.6	 & 2.5	 & 2.0	 & 2.61	 & N \\ 
NE1-1-035	 & 2014-05-14	 & 16.1	 & 9.85	 & 3.40	 & 10.42	 & 34	 & 0.1	 & 0.5	 & 0.7	 & 0.4	 & 1.3	 & 1.7	 & 2.35	 & N \\ 
NE1-1-006	 & 2014-05-14	 & 13.0	 & 9.67	 & 4.36	 & 10.61	 & 141	 & 0.4	 & 0.1	 & 2.3	 & 0.1	 & 0.7	 & 0.3	 & 2.45	 & N \\ 
NE1-1-004	 & 2014-05-14	 & 12.9	 & 10.84	 & 3.19	 & 11.30	 & 102	 & 0.4	 & 0.2	 & 1.2	 & 0.1	 & 0.9	 & 0.3	 & 2.38	 & N \\ 
E5-1-013	 & 2012-05-07	 & 15.4	 & 16.64	 & -0.27	 & 16.64	 & 34	 & 1.4	 & 0.5	 & 2.7	 & 0.4	 & 0.7	 & 1.4	 & 2.60	 & N \\

\enddata
\tablenotetext{a}{R.A. and decl. offset in projected distance from Sgr A* (R.A. offset is positive to the east).}
\tablenotetext{b}{Br$\gamma$ EW is difficult to determine due to, among other things, systematics stemming from the subtraction of local Br$\gamma$ emission. We attempt to capture this uncertainty in our errors by combining the dispersion of Br$\gamma$ EW for late-type stars at the same magnitude with the SNR-based random uncertainty.}
\end{deluxetable*}


\subsection{Spectral classification}
\label{sec:spec}

\begin{figure*}
\centering
\includegraphics[width=0.98\linewidth]{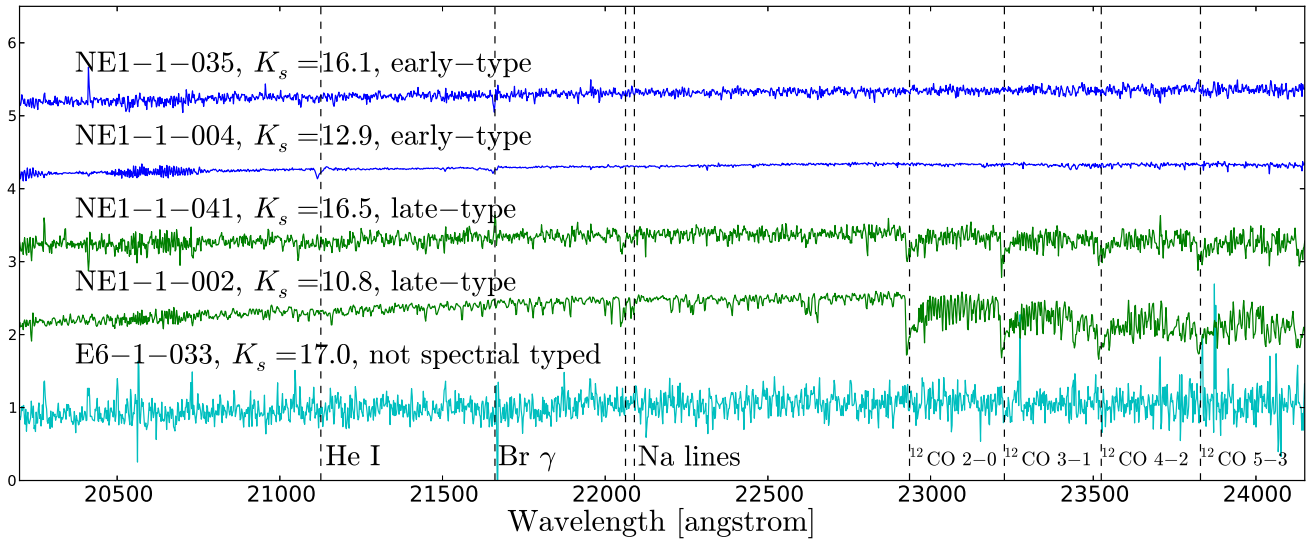}
\caption{Five typical spectra representing the types of stars seen in our sample, normalized to show the difference in absorption lines. From top to bottom; 1) A faint early-type star ($\emph{K}_s = 16.1$), the fainter of the two stars marked with blue triangles in Figure~\ref{fig:image}; 2) A bright early-type star ($\emph{K}_s = 12.9$), the brighter of the two marked with blue triangles in Figure~\ref{fig:image}; 3) A faint late-type star ($\emph{K}_s = 16.5$), the fainter of the two stars marked with large green circles in Figure~\ref{fig:image}. Notice the strong CO band head even at the low magnitude. 4) A bright ($\emph{K}_s = 10.8$) late-type star, the brighter of the two stars marked with large green circles in Figure~\ref{fig:image}; 5) A star in region E6-1 (not in Figure~\ref{fig:image}, $\emph{K}_s = 17.0$) not spectral typed due to low signal to noise ratio. Spectra of all stars from Table~\ref{tab:early},~\ref{tab:late}, and~\ref{tab:unknown} will be available in the ApJ online supplementary material.}
\label{fig:spectra}
\end{figure*}

The identification of spectral types is primarily based on the presence or absence of the CO band heads at $2.2935$, $2.3227$, $2.3525$ and $2.3830$ $\mu m$. Early-type stars are relatively featureless in the wavelength region observed, except for Br $\gamma$ $2.1661$ $\mu m$ and possibly He I $2.0581$ $\mu m$ and He I $2.1126$ $\mu m$. Detecting the Br $\gamma$ absorption is difficult due to its low equivalent width in O and B-stars (0 to 4 Angstroms), combined with systematics from background subtraction (Br $\gamma$-emitting gas is pervasive through the GC), and error induced from having to correct for the significant Br $\gamma$ absorption in the A-star we used to fix for atmospheric absorption. The He-lines are also weak, but detectable for high SNR spectra of early-type stars -- especially He I $2.1126$ $\mu m$. However, late-type stars have very strong CO band heads, which makes it an effective indicator of spectral type for almost all our sources, in addition to the Na I features at $2.2062$ $\mu m$ and $2.2090$ $\mu m$. To illustrate the difference Figure~\ref{fig:spectra} shows five spectra; a faint young star, a bright young star, a faint old star, a bright old star and a faint unclassified star. Notice that the CO band heads are strong even in the faint late-type star.

\begin{figure*}
\centering
\includegraphics[width=0.98\textwidth]{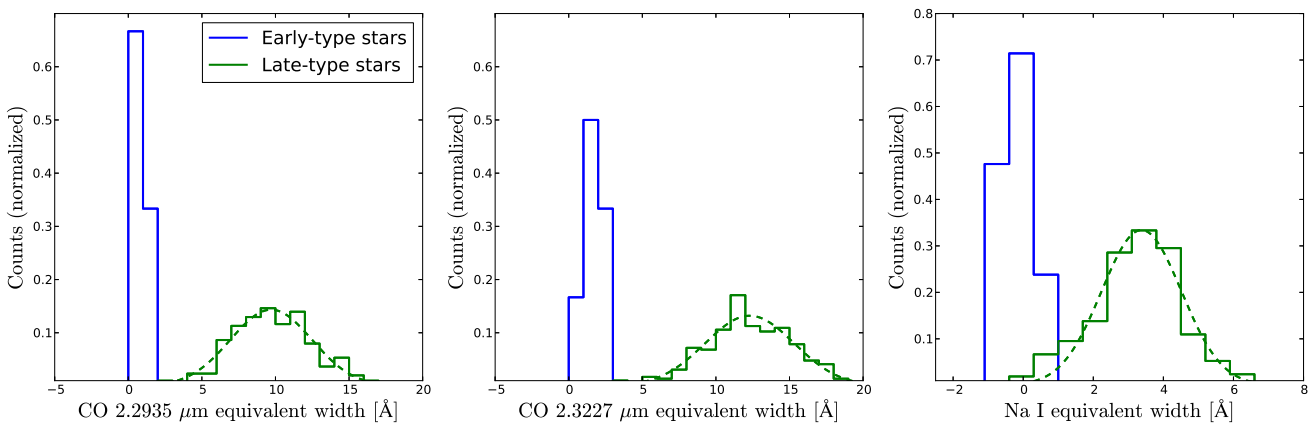}
\caption{Normalized equivalent width distributions for early-type (blue) and late-type (green) stars for CO $2.2935\mu m$ (left), CO $2.3227\mu m$ (middle) and the Na I lines (right). Best-fit Gaussian distributions are plotted for the late-type stars. Typical errors are between $0.2-1.0$\AA for the CO EWs and $0.2-0.7$\AA for the Na I lines.}
\label{fig:eqwidths}
\end{figure*}

All stars were initially classified as either early-type, late-type or unknown by visually examining their spectra. To further strengthen the spectral classifications we measured equivalent widths of CO $2.2935\mu m$, CO $2.3227\mu m$ and the Na I lines. After correcting for velocities by shifting the spectra to rest wavelengths, the equivalent widths (EWs) for all spectra with SNR\textgreater5 were found by normalizing the spectra and integrating over regions specified in \citet{forster2000} for each line. Based on these equivalent width measurements we re-evaluated a small (\textless 5) number of the initial spectral classifications. As expected we find a clear correlation between these equivalent widths and the classification of stars as late-type or early-type -- see the distributions in Figure~\ref{fig:eqwidths}. The equivalent widths are shown in Tables~\ref{tab:early} and \ref{tab:late}.

\section{Completeness and number counts}
\label{sec:comp}

\subsection{Spectroscopic Completeness}

To study the projected surface-density profiles we need to correct for both spectroscopic and photometric incompleteness. The spectroscopic completeness is, simply put, the success rate at which we are able to \emph{spectrally type} stars. More precisely it is defined as the number of stars spectrally classified over the total stars detected per magnitude bin, and is a measure of how brightness impacts our ability to spectral type stars. When there is too much noise we can no longer classify the stars as early-type or late-type, which lowers the spectroscopic completeness.

Our spectral analysis relies heavily on the presence of the CO band heads in late-type stars (and their absence in early-type stars). This naturally makes it more difficult for us to classify early-type stars than late-type stars. However, finding the spectroscopic completeness purely for early-type stars is challenging due to the low number of young stars, so we use the same spectroscopic completeness for both late-type and early-type stars. To minimize the negative effect of this in our analysis, and to increase the significance of our results, we introduce a magnitude threshold at $\emph{K}_s=15.5$. Any star fainter than this is not used for analysis. The spectroscopic completeness alone for such bright magnitudes is nearly 100\% -- we successfully spectral type all but one star brighter than the magnitude threshold. The one star that could not be spectral typed is too close to a very bright late-type star for reliable classification, and was found when we performed star planting simulations to confirm our ability to distinguish early and late-type stars near bright sources (see Appendix~\ref{App:close}). Hence, while the approximation that the early-type and late-type spectroscopic completenesses are equal is clearly not ideal, it is acceptable for the range of magnitudes used in the analysis. We also still have a high \emph{total} completeness at this threshold (74\%), and it is also the same magnitude threshold other publications have used, making comparisons easier \citep{bartko2010,do2013}.

In total 349 stars were detected, including those fainter than the magnitude threshold of $\emph{K}_s = 15.5$.\footnote{This does not include the 26 stars in the fields not used for analysis -- from fields N2-1 and N2-2 -- which brings the total up to 375 stars. 24 of these 26 stars are late-type and 2 could not be spectral typed.} Four stars were found to be foreground stars due to their very blue \emph{H-K} band colors. Of the remaining 345, we found 6 early-type stars, 286 late-type stars, and 53 stars that could not be spectral typed due to low signal-to-noise ratios.\footnote{In total 375 stars -- 310 late-type, 6 early-type and 55 unknown -- when including the regions not used for analysis.} Of the 148 stars brighter than the magnitude threshold 140 are late-type, four are early-type, three are foreground stars and only one is spectrally unidentified. See Figure~\ref{fig:totcomp} for the average spectroscopic completeness over all fields. Note that all completeness-correction is done before star magnitudes are extinction-corrected.

\subsection{Photometric Completeness}
\label{sec:photocomp}

Photometric completeness is defined as the probability of detecting a star of a certain magnitude in the data cubes, and is a measure of how well we are able to \emph{detect} stars at any given magnitude. To measure this we added artificial stars at random positions and proceeded to use the same \emph{Starfinder} routine we used to initially detect stars to determine the probability of re-detecting the artifical stars. For every image (each one corresponding to a region in Figure~\ref{fig:starlocs}) $\sim750$ stars were added per $\Delta \emph{K}_s = 0.5$ magnitude bin. The stars were re-detected by cross-referencing position and expected brightness of the added artificial star to the positions and brightnesses detected by \emph{Starfinder}. Unlike the science exposures we do not manually check these simulation cubes for obvious omissions and false detections. Because we match the detections with both positions and brightnesses it is unlikely that any false detections match up exactly with the simulated stars (ensuring we do not overestimate the photometric completeness). However, as obvious omissions are not added, it is possible that the simulations slightly underestimate the number of detected stars, and thus that the actual photometric completeness is slightly higher than what we calculate. To elaborate -- for the science exposures some stars were manually added to the list of stars after the \emph{Starfinder} routine, but due to the manual nature of this process it could not be done for the automatic completeness calculation. This leads us to detect slightly fewer stars in the simulations, which leads to a artificially low completeness. However, the number of such obvious omissions manually added in the science exposures is low, particularly over the magnitude threshold (less than one per field). It follows that the number discrepancy of stars detected between the simulations and science data is low, and so is the effect on the completeness. This has negligible effect on our completeness analysis.

\begin{figure}
\centering
  \includegraphics[width=0.8\linewidth]{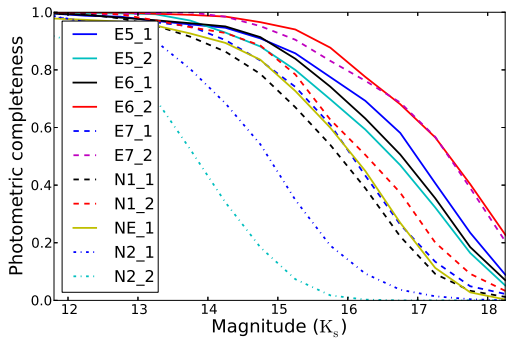}
  \caption{Photometric completeness for individual regions before extinction-correction. Regions observed with a laser guide-star for adaptive optics correction are marked as solid lines, while those observed with a natural guide-star are dotted. The two fields with lowest completeness, N2-1 and N2-2, were not used in analysis. They are plotted with dash-dot lines, but were observed with a natural guide-star for adaptive optics correction.}
  \label{fig:ngs}
\end{figure}

Having the photometric completeness for each field (Figure~\ref{fig:ngs}) we note that the observations with laser guide star adaptive optics correction obtained better results. The natural explanation to this is that the natural guide star is somewhat fainter and further away than the laser. However, it should also be noted that most of the fields observed with laser guide star are situated further from Sgr A*, so stellar density might be partly responsible for the observed completeness difference. A counter-argument to this is that a similar effect is observed in the spatial full-width half maximum of the fields (see Table~\ref{tab:obs}).

\subsection{Total Completeness}

\begin{figure}
\centering
  \includegraphics[width=0.8\linewidth]{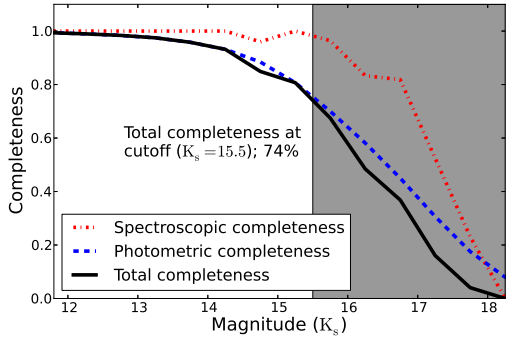}
  \caption{Completeness averaged over all fields as a function of brightness. Total completeness (black) is the multiplication of the spectroscopic (red) and photometric (blue) completeness. The region beyond the adopted magnitude threshold of $\emph{K}_s = 15.5$ is shaded grey.}
  \label{fig:totcomp}
\end{figure}

Using the spectroscopic completeness and average photometric completeness we can find an overall completeness curve. This total completeness curve measures the average likelihood of a detection and classification of any magnitude star in any our fields, i.e. the overall completeness over all fields. At the magnitude threshold $\emph{K}_s=15.5$ the total completeness varies between 59-92\% for the analyzed fields.  At any magnitude brighter than this threshold the total completeness is dominated by the photometric completeness rather than the spectroscopic completeness. In other words, if the star can be detected at all and is brighter than $\emph{K}_s=15.5$, then it can almost certainly be spectral typed. The major limiting factor on our completeness is source confusion due to low spatial resolution. Across all the fields, the average total, photometric, and spectroscopic completenesses at the magnitude threshold are 74\%, 75\%, and 98\% respectively. The total completeness curve, together with the spectroscopic and photometric completeness curves, can be seen in Figure~\ref{fig:totcomp}.


\begin{figure}
\centering
\includegraphics[width=1.0\linewidth]{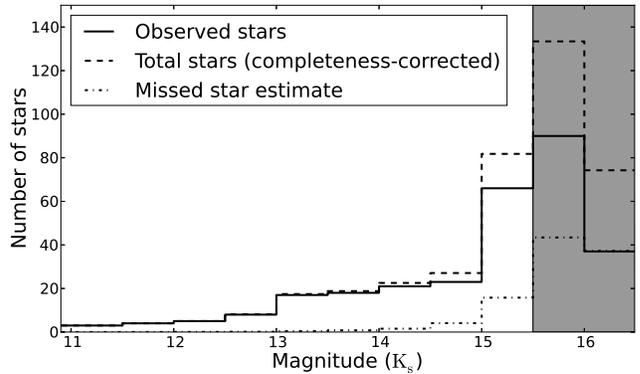}
\caption{$\emph{K}_s$-band number counts of all observed stars, with and without completeness-correction. The number of stars the completeness-correction adds is also plotted. The drop in total number count beyond the magnitude cutoff is likely due to large errors on completeness at fainter magnitudes. These counts have been extinction-corrected to an average value of $A_{\emph{K}_S} = 2.7$.}
\label{fig:missed}
\end{figure}

With these completeness estimates we can approximate the total number of stars we could not detect or classify due to photometric or spectral incompleteness. Completeness-correction over all regions adds 22.8 late-type stars and only 0.3 early-type stars.  The total number count of stars with and without completeness-correction is shown in Figure~\ref{fig:missed}.

\section{Results}
\label{sec:res}

\subsection{Projected surface-density profiles}
\label{sec:mydens}

\begin{figure}
\centering
\includegraphics[width=0.98\linewidth]{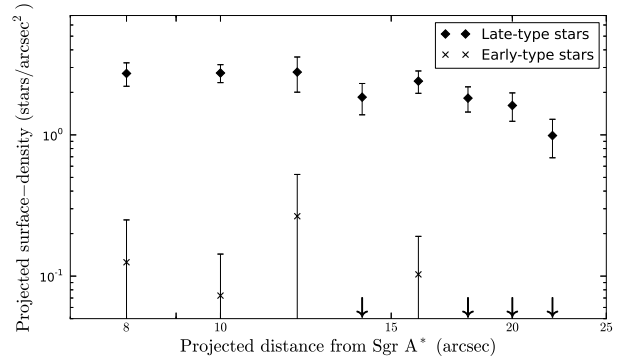}
\caption{Extinction-corrected, completeness-corrected projected surface-density of early-type (crosses) and late-type (diamonds) stars. 139 late-type and five early-type stars brighter than the magnitude threshold were detected, with 22.8 and 0.3 respectively added by completeness correction. Bin size is two arcseconds, and $1\sigma$ errors are indicated. Downward arrows represent non-detections and do not signify bin upper limits. Spatial coverage is 8.03, 14.20, 3.87, 7.81, 11.33, 12.14, 10.90, and 9.99 arcsec$^2$ for each bin respectively. No early-type stars were detected further than $17\arcsec$ away from Sgr A*.}
\label{fig:dens}
\end{figure}

Figure~\ref{fig:dens} shows the completeness-corrected projected surface-density profile for early- and late-type stars $\emph{K}_s \leq 15.5$. Four early-type stars brighter than the magnitude threshold were detected. Three of these four stars were detected in the northern regions; two in NE1-1 and one in N1-1. There is a significant dearth of early-type stars in the outer regions (see Section~\ref{sec:comparison} and \ref{sec:dis}). The most distant star from Sgr A* we detect is at a projected distance of $16.6\arcsec$, and is also the only early-type star in the eastern fields. Its magnitude indicates that it is a B-star, and it has a radial velocity opposite of what one would expect from a member of the clockwise young stellar disk.

\subsection{Comparison to other publications}
\label{sec:comparison}

\begin{figure}
\centering
  \includegraphics[width=\linewidth]{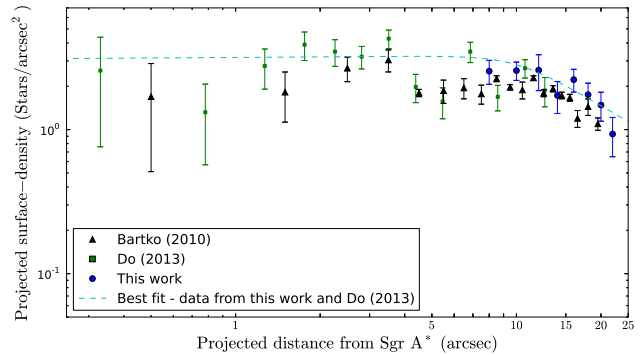}
  \caption{Extinction- and completeness-corrected projected surface-density of late-type stars from this work (blue circles), \citet{do2013} (green squares) and \citet{bartko2010} (black triangles). Bin size is $2\arcsec$ for data from this publication. A dotted line representing the best fit broken power-law, found using only the data from \citet{do2013} and this publication, is plotted in cyan.}
  \label{fig:latecomp}
\end{figure}

\begin{figure}[h]
\begin{minipage}[b]{\linewidth}
  \centering
  \includegraphics[width=1.0\linewidth]{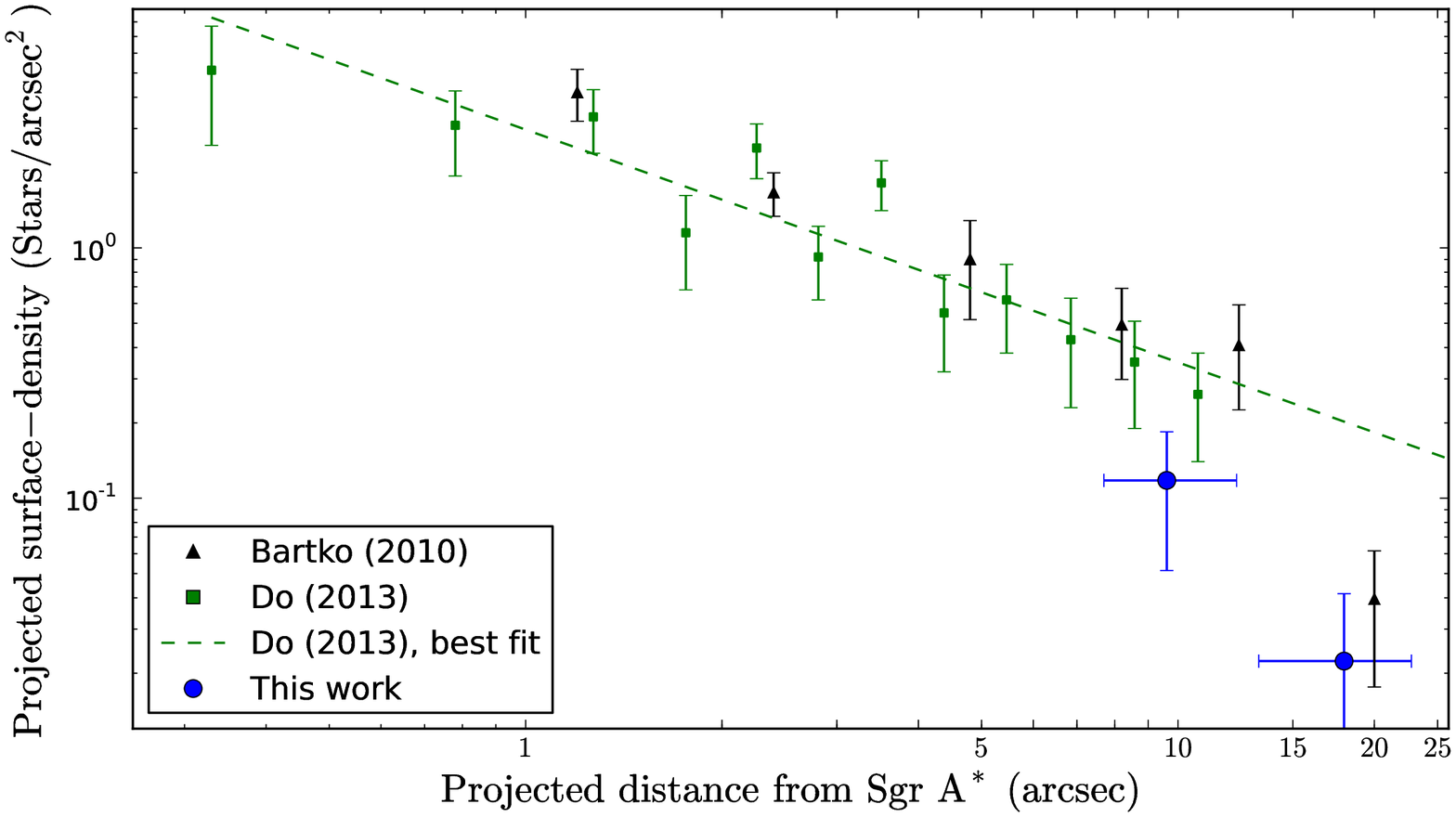}
\end{minipage}
\begin{minipage}[b]{\linewidth}
  \centering
  \includegraphics[width=1.0\linewidth]{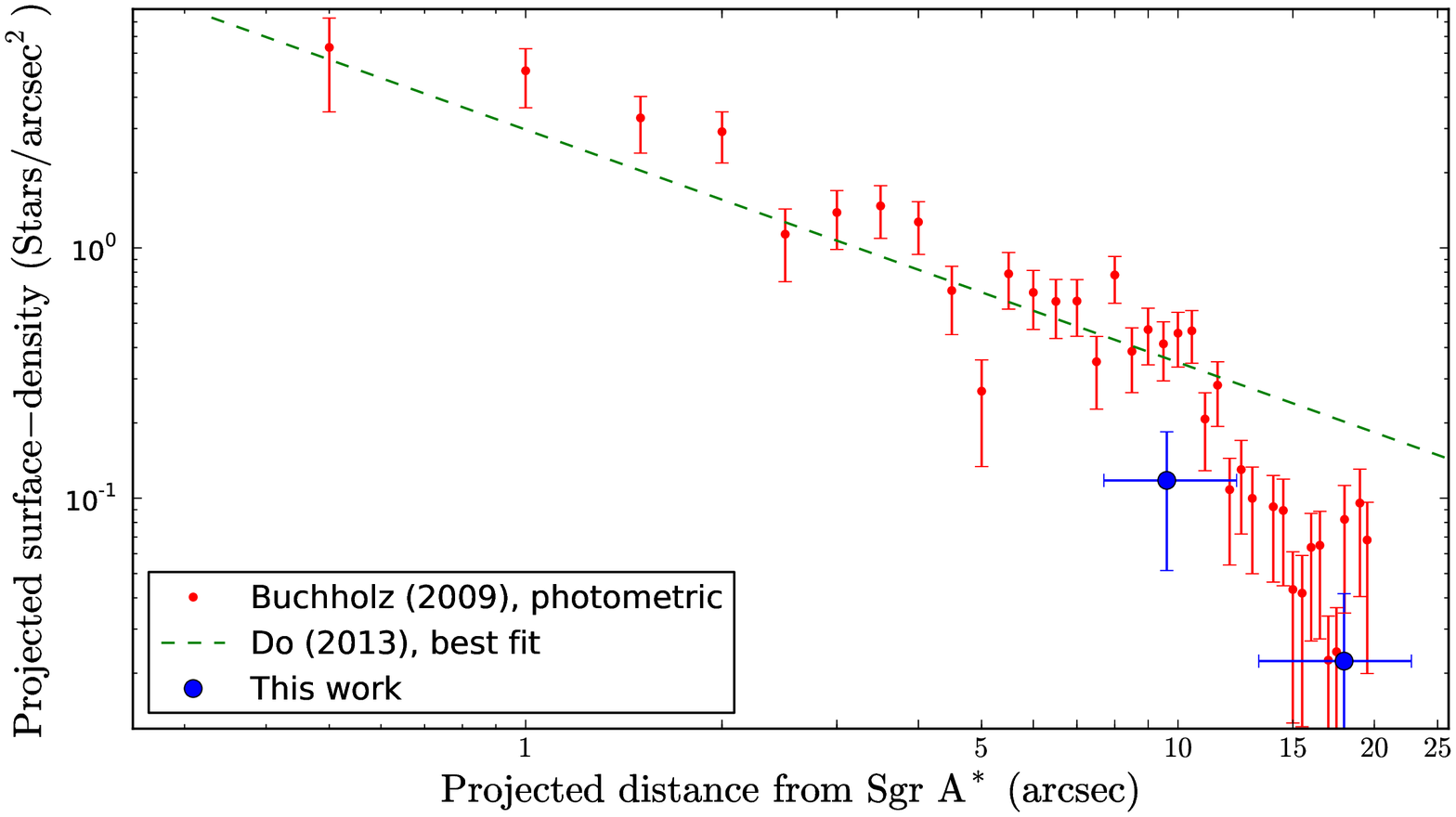}
  \caption{Extinction- and completeness-corrected projected surface-density of early-type stars. The first figure compares the results from this paper (blue circles) to the spectroscopic results from \citet{do2013} (green squares) and \citet{bartko2010} (black triangles). The second figure compares the same results from this paper (blue circles) to the photometric results of \citet{buchholz2009} (smaller red circles). The most probable fit from \citet{do2013} is also plotted in both figures, with $\Gamma = 0.93$. The two data points from this publication are measured separately, one from the northern fields ($9.\arcsec6$) and one from the eastern fields ($18.\arcsec0$). The two bins have ranges of ($7.\arcsec7$--$12.\arcsec3$) and ($13.\arcsec3$--$22.\arcsec8$) -- these ranges are shown as x-errorbars. The stellar densities in the two observed regions are $0.12 \pm 0.07$ and $0.022 \pm 0.019$ stars arcsec$^{-1}$, respectively. These data points are equal to those of Figure~\ref{fig:dens} with a different binning. The points from \citet{bartko2010} represents their combined sample of all young stars.}
  \label{fig:earlcomp}
\end{minipage}
\end{figure}

Other publications, most notably \citet{bartko2010} and \citet{do2013}, have also measured projected surface-density profiles for the Galactic center. Using the data from these publications we again plot late-type and early-type density profiles in Figure~\ref{fig:latecomp} and~\ref{fig:earlcomp}.

The late-type surface-density can be seen in Figure~\ref{fig:latecomp} and is consistent with what we expect from previous surveys. This work finds evidence for a broken power-law late-type surface-density profile, which is in agreement with the findings of \citet{buchholz2009} and \citet{do2013}. To fit the broken power-law we use the Bayesian parameter estimation method that does not require binning the data, introduced in \citet{do2013B}, and the following profile:

\begin{equation}
\rho_{\star}(R) = A (\frac{R}{R_b})^{-\gamma}(1+(R/R_b)^{\delta})^{(\gamma-\alpha)/\delta}
\end{equation}

Where $\gamma$ is the inner power law slope, $\alpha$ is the outer slope, $R$ is the projected radius from Sgr A*, $R_b$ is the break radius, $A$ is a normalization factor and $\delta$ is the sharpness of the transition between the two slopes. We use the data from this publication and in \citet{do2013} to find the best fit, which has a break radius of $10\arcsec \pm 5\arcsec$ and slopes $\gamma = -0.013 \pm 0.170$ and $\alpha = 1.15 \pm 0.52$. The normalization factor is $3.27 \pm 0.47$ stars arcsec$^{-2}$, and $\delta$ is $5.44 \pm 3.45$. The fitted line is shown in cyan in Figure~\ref{fig:latecomp}.

The early-type projected surface-density can be seen in Figure~\ref{fig:earlcomp}. The data from this paper has been divided into two bins, one for the northern regions (at $9.\arcsec6$) and one for the eastern regions (at $18.\arcsec0$). It is compared first to the spectroscopic data from \citet{do2013} and \citet{bartko2010}, and second to the photometric data from \citet{buchholz2009}. There is a clear drop-off in stellar density below that predicted by the single power law from \citet{do2013} outside $13\arcsec$, a drop-off that is observed in our data, the data from \citet{bartko2010}, and in the photometric data from \citet{buchholz2009}. The data from \citet{buchholz2009}, despite being non-spectroscopic and prone to false detections, also seem to match the overall trend of the spectroscopic observations well. From \citet{do2013} we expect the early-type stars to follow a best fit power law of $\Gamma = 0.90 \pm 0.09$, which corresponds to an expected $9.1$ young stars outside $12\arcsec$ in our data. Only a single young star is detected in the region, with completeness-correction adding 0.16 (or at most one) more. There is also a single non-classified star in the region, which could not be classified due to proximity to a brighter source. In contrast to sources with low SNR spectra it is no more likely to be early-type than any other star (see Appendix~\ref{App:close}). Even so, if we assume that this star is actually early-type or that we missed one for a total of two early-type stars in the region, we observe at least seven less early-type stars than expected, which corresponds to a Poisson probability of 0.57\% and a $2.5\sigma$ deficit. If we make the assumption that we detected all the young stars our result corresponds to a 0.11\% Poisson probability and a $3\sigma$ deficit. We conclude, based on the large area sampled, that there is a significant lack of young stars further than $13\arcsec$ away from Sgr A*, possibly indicating an edge in the young stellar cluster somewhere between $12\arcsec$ and $14\arcsec$ ($\sim0.50$pc). The last data point in the projected surface-density profile of young stars from \citet{bartko2010} strengthens this conclusion. It also indicates that the edge has some azimuthal symmetry, as the Bartko data outside $12\arcsec$ is mostly situated north of Sgr A* (compared to our fields situated east of Sgr A*), and both follow the general trend of \citet{buchholz2009}.

The surface-density in the northern fields -- the data point at $9.\arcsec6$ -- is lower than the data from \citet{bartko2010} and \citet{do2013}. When comparing to overlapping regions from \citet{bartko2010} we note that the low number of early-type stars in the northern fields is consistent for the overlapping sections. The northern fields only have half the area of the eastern fields, leading to a low statistical significance, and the results are consistent to within $1\sigma$ when compared to \citet{do2013}.

\section{Discussion}
\label{sec:dis}

\subsection{The detection of an outer edge, the infalling cluster scenario and the clockwise disk}

Previous publications have shown that the observed density profile of the GC young stellar cluster is consistent with a single power-law inside $12\arcsec$ \citep{paumard2006,do2009,lu2009,bartko2010,do2013,yelda2014}. Our observations demonstrate that extending this power-law to larger radii overpredicts the number of young stars with a significance of at least $2.5\sigma$. We thus conclude that there is a drop in the surface-density of the young stellar cluster, possibly indicating an outer edge to the young stellar cluster in the Galactic center, at $\sim13\arcsec$ ($0.52$ pc). This strengthens the validity of the results from \citet{buchholz2009}. The spectroscopic nature of our observations removes the bias from photometric surveys like that of \citet{buchholz2009} -- our result is thus more definite. The observed stellar density profile is consistent with a broken power law. N-body simulations have recently shown that relaxation times of disk stars can produce deviations from a single power-law in times comparable to the estimated age of the disk (6 Myr - \citet{subr2014}), which makes such a broken power-law structure intriguing. 

The detection of an outer edge to the young stellar cluster shows that the young stellar disk is unlikely to have been the result of an infalling high-density cluster core. Inward movement and subsequent tidal disruption of a cluster from a larger radius has previously been considered a possibility for the origin of the young stellar disk \citep{gerhard2001}. However, recent results have favored in-situ formation in a massive accretion disk \citep[][others]{paumard2006, lu2009, bartko2010}. Our results add to the evidence against the infalling cluster hypothesis. Such a scenario would result in a shallow density profile, on the order of $r^{-0.75}$ \citep{berukoff2006}. This comes with the caveat that mass segregation could cause the observed density profile for the massive O- and B-stars -- the only stars that are bright enough to be detected with current instruments -- to be steeper than the actual density profile for the cluster. However, the detection of an edge (or at least a significant decrease in the surface-density) at around $12-14\arcsec$ is another strong argument for \emph{in situ} star formation, as such an edge would simply mark the radius from Sgr A* at which stars could no longer form. An infalling cluster would leave a trail of stars, which we do not observe in this study.

It also follows that our results strengthens the hypothesis that the scale of the recent (4-6Myr) star formation event in the Galactic center was within 0.5pc of Sgr A* \citep{bartko2010}. Recent publications have often more or less assumed a similar scale \citep{paumard2006, subr2014}; we provide evidence that this assumption is consistent with observed data.

Our results show an edge to the young stellar cluster in the eastern direction from Sgr A*, and combined with results from \citet{bartko2010} they also imply a similar edge in the northern direction. It is currently unclear if this cluster edge also represents the outer edge of the clockwise disk of young stars \citep{paumard2006,lu2009,do2009,bartko2009,bartko2010,do2013,yelda2014}. If the disk extends indefinitely with no warp it is projected to lie on top of the regions we observed east of Sgr A*, i.e. our observations outside $13\arcsec$. We do not detect any disk stars in this region, which strongly indicates that either (1) the disk has an outer edge closer to Sgr A* than $13\arcsec$ or (2) the disk is warped. Several prior publications \citep{bartko2009, kocsis2011} have suggested such a warp of the disk, which could mean that our observations are off the disk plane. Regardless, a disk edge outside the cluster edge seems difficult to reconcile with observations and theory, particularly as results from \citet{yelda2014} show a decreasing statistical significance of the disk with distance from Sgr A*. For now it seems likely that the young stellar disk -- with or without warp -- has an edge at $\sim13\arcsec$, or possibly closer to Sgr A*.


\subsection{Origin of the cluster edge: gas disk instability}

Here we explore the possibility that the truncation in the stellar disk may represent the radius of gravitational instability in the original gas disk,  which would explain the observed edge to the young stellar disk and cluster. Star formation in a gaseous disk requires it to be gravitationally unstable in order to collapse. This instability will take place if the gas pressure and rotational support is less than the local self-gravity of the disk, and is represented by the Toomre $Q$ parameter \citep{toomre1964}:
\begin{equation}
Q = \frac{v_{Kep} \sigma}{G \pi r \Sigma_{gas}},
\end{equation}
where $v_{Kep}$ is the local Keplerian velocity, $\sigma$ is the velocity dispersion of the gas, $r$ is the distance from the black hole, and $\Sigma_{gas}$ is the surface-density of the gas. $Q < 1$ indicates that the gas is gravitationally unstable and will collapse. The disk is stable for $Q > 1$. Prior estimates of disk instabilities at the Galactic center disk have also used the Toomre Q parameter for analysis \citep[e.g.,][]{nayakshin2007, levin2007}.

In order to estimate Q, we need to make a few assumptions about the original gas disk. We assume that the gas disk is in Keplerian orbit around the black hole, because the disk is well within the its gravitational radius of influence.The gas velocity dispersion is unknown, but we can estimate it with two approaches: (1) using a fit to the observed velocity dispersion profile of young stars, or (2) using a velocity dispersion that is constant with disk radius. The first assumption yields $\sigma_{gas,r}(r) \propto 1/\sqrt{r}$ using data from \citet{yelda2014}, while the second assumption has $\sigma_{gas,k}(r) = $ constant.  We assume the gas disk surface-density $\Sigma$ follows the currently observed stellar surface-density profile $\Sigma_{stars}(r) \propto r^{-1.7}$ \citep{paumard2006,lu2009,bartko2010,yelda2014}. To obtain the absolute values of $\Sigma_{gas}$ requires knowledge of the star formation efficiency as well as the current total disk mass. Because these parameters, especially the star formation efficiency, are poorly constrained, the absolute value of Q will also be highly uncertain. Here, we take instead the approach of calculating the radial dependence of Q to determine whether the edge represents an increasing Q value with radius. For $\sigma_{gas,r}$ the radial dependence of Q is:
\begin{equation}
Q(r) \propto \frac{r^{-0.5} r^{-0.5}}{r r^{-1.7}} \propto r^{-0.3},
\end{equation}
while for a constant $\sigma_{gas,k}$:
\begin{equation}
Q(r) \propto \frac{r^{-0.5}}{r r^{-1.7}} \propto r^{0.2},
\end{equation}
The two assumptions about $\sigma_{gas}$ lead to different interpretations of the disk instability. For a radially dependent $\sigma_{gas}$, Q decreases with radius, which would suggest that the disk instability criteria cannot explain the outer disk edge. For a constant $\sigma_{gas}$, the interpretation would be the reverse. With Q increasing slightly with radius, this would be consistent with the disk having an outer edge because it is stable to gravitational collapse beyond a certain radius. 

Our simple analysis of the disk instability criterion above shows that given the lack of knowledge of the initial conditions of the gas, it is inconclusive whether disk instability should lead to an observed edge to the currently observed stellar disk. The main emphasis of this section is to illuminate the need for more research on the initial conditions of the gas disk. Our calculations have large uncertainties, but with more precise initial parameters a similar procedure could lead to meaningful conclusions on both the history and the current state of the young stellar cluster.

\subsection{Cloud-cloud collision}

Simulations of cloud-cloud collisions, such as those by \citet{hobbs2009}, shows star formation that is qualitatively consistent with our results. These simulations result in an inner stellar disk of young stars along with streams of stars outside the plane of the disk. While the region where stars formed in \citep{hobbs2009} is consistent with our present observations, the authors do not explore the different physical parameters of the initial clouds that would affect the size of the resulting cluster. More theoretical work is needed to constrain the radial range where cloud conditions were sufficient to form stars. Our observation of a scale radius of $\sim0.5$ pc may help constrain the radius at which the cloud-cloud collision occurred. In addition, this scale may provide tests for physical mechanisms that limit star formation in this region, such as feedback both from star formation and from accretion onto the supermassive black hole.


\section{Conclusion}
\label{sec:conc}

We report the results of a new spectroscopic survey of the Galactic center with a radial extent of  $7\arcsec$ to $23\arcsec$ (0.28-0.92 pc). We find a total of 349 stars in the regions used for analysis, 148 being brighter than the magnitude threshold at $\emph{K}_s = 15.5$, a magnitude at which we are able to spectral type stars with a 74\% completeness. We detect four early-type (young) stars brighter than the magnitude threshold at $\emph{K}_S = 15.5$, with an additional two fainter young stars. Only one of these six young stars is located further from Sgr A* than $12\arcsec$, despite our expectation of nine young stars brighter than the magnitude threshold in the region, which strongly indicates a sharp surface-density drop-off of the young stellar cluster at $\sim 13\arcsec$ (0.52 pc) in projected distance from Sgr A*.

This drop-off of the early-type projected surface-density is found to be consistent with a broken power-law fit, which is inconsistent with the theory that the young stars originated from a tightly bound infalling stellar cluster. The \emph{in situ} theory is deemed more credible. However, we find no simple answer as to why this type of star formation would cause an outer disk edge at $13\arcsec$. We note that a simple disk instability criterion could possibly explain the disk edge, but we are unable to make any conclusive remarks due to a lack of knowledge of the initial gas disk conditions. Similarly, we note that a cloud-cloud collision could produce the observed distribution of young stars, but we currently lack the theoretical framework to further explore the idea. The surface-density of late-type stars is also measured, and is found to match well both with spectroscopic surveys closer to Sgr A* as well as imaging surveys with a larger radial and azimuthal extent. 

The authors thank the staff of the Gemini North observatory for all their help in obtaining the new observations, and Rainer Sch\"odel for providing us with accurate magnitudes and an extinction map of the Galactic center. This paper uses data from Gemini observations GN-2012A-Q-41 and GN-2014A-Q-71, together with HST programs GO-12182 and GO-11671. We gratefully acknowledge the support of the Dunlap Institute for Astronomy \& Astrophysics, University of Toronto. A.M.G. is supported by the NSF grants AST-0909218 and AST-1412615, and the Lauren Leichtman \& Arthur Levine Chair in Astrophysics

\emph{Facilities:} Gemini North (NIFS)

\appendix
\section{\\Close companions: star planting simulations} \label{App:close}

The photometric and spectroscopic completeness described in Section~\ref{sec:comp} measure whether a star is detected and whether the extracted spectra can be classified as early-type or late-type. We also need to ensure that the stars we manage to spectral type are spectral typed correctly -- this is not included in the initial completeness. Thus we need to ensure that no spectrum is extracted and mistakenly classified.

In order to determine the impact of nearby sources on our ability to determine spectral types, we performed star planting simulations with close pairs of stars. Note that these are not binary stars, which is a different source of error, but rather stars that are close to each other in pixel distance in our data.\footnote{In binary star systems the brighter star usually dominates the flux output, which presumably decreases the effect of binaries on our data.} Any relatively consistent background emission should be removed by the background subtraction and would have little impact, particularly in the bright stars used for analysis ($\emph{K}_s \leq 15.5$). However, the background subtraction does a poor job of subtracting excess flux from nearby stars. In other words, our spectral typing is robust for solitary sources, but there has to be a check for contamination in the spectra of close companions (when stars are less than five spatial pixels away from each other). This is especially troublesome as the equivalent widths of the CO band heads vary between late-type stars, which makes it difficult to distinguish between a late-type star with low CO equivalent widths and an early-type star with contamination from a late-type star.

\begin{figure}
\centering
\includegraphics[width=3.3in]{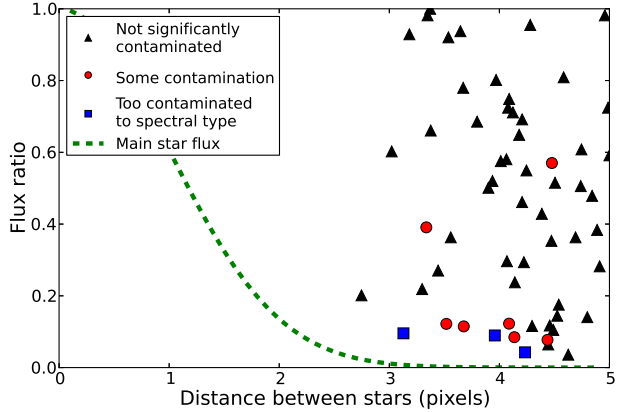}
\caption{A plot of close companions in terms of the distance between stars and their respective flux ratio. Only pairs for which both stars were initally spectral typed are included. Stars can be plotted more than once in different pairs. Black triangles are pairs for which contamination was deemed negligible after star planting simulations, red squares are pairs where contamination is significant but spectral type is still known, while blue circles are pairs for which the spectral typing of the faint star was changed to unknown due to significant contamination. The red circle at 4.5 pixel distance with a large flux ratio is due to an especially crowded stellar region.}
\label{fig:close}
\end{figure}

To identify which pairs of stars would need to be examined further a plot of distance between stars versus flux ratio of the stars was created (pixel distance versus $\frac{F_{faint}}{F_{bright}}$ -- see Figure~\ref{fig:close}). A line symbolizing the flux of the main star was plotted, assuming a FWHM of 1.5 pixels for every field.\footnote{As FWHM varies between fields this is an approximation, but is close enough to get a rough idea of which stars should be further examined.} Star pairs with either a) a seperation below five pixels or b) a seperation below seven pixels and a flux ratio below 0.1 were classified as candidates for star planting simulations to manually check for contamination.

These star planting simulations take into account that all close companions are initially identified as late-type stars. There are no stars classified as early-type that are close enough to other stars to be at risk. Hence the objective of the simulations is to ensure that the spectrum of an early-type star cannot be interpreted as the spectrum of a late-type star if it is close to a brighter late-type star. 

For each pair of late-type close companions we ran 25 simulations where faint stars with early-type spectra were placed at the same distance and flux ratio to the brighter star. The early-type spectra were from the Kurucz models, convolved to fit our spectral resolution and with added noise such that signal-to-noise is proportional to the stellar flux. The simulated stars were placed around the least populated areas, but always at the same distance as the original faint star (these positions were usually on the opposite side of the original faint star unless the companions were close to the edge of the data cube). The spectra were then extracted and inspected individually, being compared to the original faint star spectrum. For each simulation we determined whether the spectrum showed CO lines and could be mistaken as a late-type star. If the simulated star was spectral typed as late-type for more than two of the 25 spectra (over 10\%) the original faint companion is listed in Table~\ref{tab:late} as contaminated, while if all the simulations could be spectral typed as late-type the spectral type was changed to unknown (as with E7-1-009).

In general contamination was only a problem for pairs with a very low flux ratios where both stars already had a faint magnitude (the fainter of the two was almost always fainter than the magnitude threshold). This is unsurprising -- low flux ratios where the brighter star is not one of the brightest stars in the region implies a very faint companion with a low signal-to-noise. Such stellar spectra are clearly suspect to contamination. Differing spatial resolution between fields and stellar density were also found to factor in which stars were contaminated. Only a single spectrum from a star brighter than the magnitude threshold ($\emph{K}_s = 15.5$) is contaminated (E7-1-009, to such a degree that it was re-classified as unknown spectral type). Three stars were re-classified as unknown spectral type after these simulations (E7-1-009, E5-2-024, and E6-2-011), while a further six were classified as contaminated -- see the rightmost column in Table~\ref{tab:late}.  The distribution of contaminated stars related to distance and flux ratio can be seen in Figure~\ref{fig:close}.\newline

\bibliography{refs}

\section{Tables: Late-type and unknown stars} \label{App:tables}
\LongTables
\begin{deluxetable*}{lccccccccccccrr}
\tabletypesize{\scriptsize}
\tablecolumns{15}
\tablewidth{0pc}
\tablecaption{NIFS Observations of Late-Type Stars
\label{tab:late}}
\tablehead{    
  \vspace{-0.12cm} &  &  &
  \colhead{$\Delta$R.A.\tablenotemark{a}} &
  \colhead{$\Delta$Decl\tablenotemark{a}} &
  \colhead{R} &  &
  \colhead{CO 2.2935} &
  \colhead{Err} &
  \colhead{CO 2.3227} &
  \colhead{Err} &
  \colhead{Na I} &
  \colhead{Err} &  &  \\
  \colhead{Name} \vspace{-0.15cm} &
  \colhead{Date} &
  \colhead{$\emph{K}_s$} &   &  &  &
  \colhead{SNR} &  &  &  &  &  &  &
  \colhead{$A_{\emph{K}_s}$} &
  \colhead{Contam}    \\
  &  &  &
  \colhead{($\arcsec$)} &
  \colhead{($\arcsec$)} &
  \colhead{($\arcsec$)} &
  &
  \colhead{EW (\AA)} &
  \colhead{(\AA)} &
  \colhead{EW (\AA)} &
  \colhead{(\AA)} &
  \colhead{EW (\AA)} &
  \colhead{(\AA)} &  &
}
\startdata

N1-2-030	 & 2012-05-11	 & 15.8	 & 4.27	 & 6.48	 & 7.76	 & 20	 & 10.1	 & 0.9	 & 11.1	 & 0.6	 & 4.1	 & 0.5	 & 2.67	 & Y \\ 
N1-1-022	 & 2012-05-11	 & 15.6	 & 2.28	 & 7.45	 & 7.79	 & 32	 & 9.7	 & 0.5	 & 10.7	 & 0.4	 & 2.7	 & 0.4	 & 2.63	 & N \\ 
N1-2-019	 & 2012-05-11	 & 15.8	 & 4.21	 & 6.64	 & 7.86	 & 25	 & 10.1	 & 0.6	 & 12.7	 & 0.4	 & 3.4	 & 0.4	 & 2.67	 & N \\ 
N1-2-010	 & 2012-05-11	 & 15.6	 & 4.74	 & 6.35	 & 7.92	 & 29	 & 11.3	 & 0.5	 & 13.1	 & 0.4	 & 3.5	 & 0.4	 & 2.75	 & N \\ 
N1-1-045	 & 2012-05-11	 & 12.3	 & 3.64	 & 7.22	 & 8.08	 & \textgreater40	 & 6.8	 & 0.2	 & 5.6	 & 0.2	 & 0.5	 & 0.1	 & 2.56	 & N \\ 
N1-1-039	 & 2012-05-11	 & 15.9	 & 3.44	 & 7.33	 & 8.10	 & 16	 & 4.3	 & 1.0	 & 5.7	 & 0.7	 & 1.4	 & 0.8	 & 2.50	 & N \\ 
N1-1-027	 & 2012-05-11	 & 15.8	 & 0.95	 & 8.11	 & 8.16	 & 13	 & 7.4	 & 1.1	 & 12.3	 & 0.6	 & 0.4	 & 0.8	 & 2.62	 & N \\ 
N1-2-028	 & 2012-05-11	 & 15.9	 & 4.91	 & 6.58	 & 8.21	 & 25	 & 9.5	 & 0.6	 & 10.4	 & 0.5	 & 0.7	 & 0.5	 & 2.78	 & N \\ 
N1-2-013	 & 2012-05-11	 & 15.2	 & 5.57	 & 6.04	 & 8.21	 & 24	 & 10.7	 & 0.6	 & 12.3	 & 0.5	 & 1.9	 & 0.5	 & 2.80	 & N \\ 
N1-1-028	 & 2012-05-11	 & 15.7	 & 2.62	 & 7.80	 & 8.22	 & 19	 & 8.3	 & 0.8	 & 9.7	 & 0.6	 & 1.4	 & 0.6	 & 2.64	 & N \\ 
N1-2-042	 & 2012-05-11	 & 16.4	 & 5.22	 & 6.40	 & 8.26	 & 21	 & 7.8	 & 0.8	 & 9.2	 & 0.5	 & 2.6	 & 0.6	 & 2.78	 & N \\ 
N1-1-011	 & 2012-05-11	 & 15.0	 & 1.90	 & 8.06	 & 8.28	 & 20	 & 4.2	 & 0.8	 & 4.3	 & 0.6	 & 0.9	 & 0.6	 & 2.59	 & N \\ 
N1-1-031	 & 2012-05-11	 & 15.8	 & 3.09	 & 7.73	 & 8.32	 & 19	 & 8.4	 & 0.8	 & 10.5	 & 0.6	 & 1.3	 & 0.6	 & 2.54	 & N \\ 
N1-2-015	 & 2012-05-11	 & 15.7	 & 5.45	 & 6.35	 & 8.37	 & 19	 & 9.1	 & 0.9	 & 9.2	 & 0.6	 & 4.2	 & 0.6	 & 2.77	 & N \\ 
N1-2-007	 & 2012-05-11	 & 14.5	 & 5.60	 & 6.25	 & 8.39	 & 32	 & 9.9	 & 0.5	 & 14.4	 & 0.3	 & 3.4	 & 0.4	 & 2.78	 & N \\ 
N1-1-020	 & 2012-05-11	 & 15.8	 & 1.74	 & 8.25	 & 8.44	 & 19	 & 12.4	 & 0.8	 & 12.6	 & 0.7	 & 2.4	 & 0.6	 & 2.59	 & N \\ 
N1-2-009	 & 2012-05-11	 & 14.9	 & 5.88	 & 6.09	 & 8.46	 & 34	 & 4.4	 & 0.5	 & 5.2	 & 0.4	 & 1.3	 & 0.4	 & 2.80	 & N \\ 
N1-1-002	 & 2012-05-11	 & 13.4	 & 1.40	 & 8.51	 & 8.63	 & \textgreater40	 & 4.9	 & 0.2	 & 8.9	 & 0.2	 & 1.1	 & 0.2	 & 2.61	 & N \\ 
N1-2-026	 & 2012-05-11	 & 15.6	 & 5.22	 & 6.87	 & 8.63	 & 25	 & 9.5	 & 0.7	 & 12.0	 & 0.4	 & 2.2	 & 0.6	 & 2.80	 & N \\ 
N1-1-023	 & 2012-05-11	 & 15.5	 & 2.40	 & 8.33	 & 8.67	 & 18	 & 5.7	 & 1.0	 & 10.0	 & 0.7	 & 1.9	 & 0.7	 & 2.64	 & N \\ 
N1-1-012	 & 2012-05-11	 & 15.0	 & 2.70	 & 8.25	 & 8.68	 & 26	 & 11.5	 & 0.6	 & 12.8	 & 0.5	 & 4.8	 & 0.4	 & 2.64	 & N \\ 
N1-2-029	 & 2012-05-11	 & 15.8	 & 4.58	 & 7.41	 & 8.71	 & 27	 & 8.2	 & 0.8	 & 10.4	 & 0.3	 & 1.4	 & 0.4	 & 2.64	 & N \\ 
N1-2-005	 & 2012-05-11	 & 14.3	 & 5.11	 & 7.05	 & 8.71	 & 31	 & 15.1	 & 0.4	 & 15.4	 & 0.3	 & 3.3	 & 0.3	 & 2.78	 & N \\ 
N1-2-023	 & 2012-05-11	 & 15.6	 & 6.21	 & 6.17	 & 8.76	 & 29	 & 6.1	 & 0.5	 & 9.0	 & 0.4	 & 2.6	 & 0.4	 & 2.77	 & Y \\ 
N1-1-019	 & 2012-05-11	 & 15.8	 & 1.54	 & 8.62	 & 8.76	 & 19	 & 5.8	 & 0.9	 & 8.5	 & 0.7	 & 1.1	 & 0.5	 & 2.59	 & Y \\ 
N1-1-006	 & 2012-05-11	 & 14.1	 & 2.92	 & 8.29	 & 8.79	 & 34	 & 9.8	 & 0.5	 & 13.1	 & 0.4	 & 3.6	 & 0.4	 & 2.62	 & N \\ 
N1-2-020	 & 2012-05-11	 & 15.5	 & 6.16	 & 6.33	 & 8.83	 & 19	 & 6.8	 & 1.0	 & 8.1	 & 0.5	 & 0.5	 & 0.6	 & 2.80	 & N \\ 
N1-1-014	 & 2012-05-11	 & 15.1	 & 2.59	 & 8.50	 & 8.88	 & 18	 & 7.8	 & 0.9	 & 9.9	 & 0.6	 & 2.4	 & 0.7	 & 2.65	 & N \\ 
N1-1-001	 & 2012-05-11	 & 13.3	 & 3.51	 & 8.17	 & 8.89	 & \textgreater40	 & 12.2	 & 0.4	 & 14.8	 & 0.2	 & 4.5	 & 0.3	 & 2.63	 & N \\ 
N1-2-014	 & 2012-05-11	 & 15.5	 & 4.60	 & 7.61	 & 8.89	 & 23	 & 13.2	 & 0.7	 & 15.5	 & 0.4	 & 5.1	 & 0.4	 & 2.64	 & N \\ 
N1-1-016	 & 2012-05-11	 & 15.2	 & 2.30	 & 8.63	 & 8.93	 & 19	 & 6.1	 & 0.9	 & 10.9	 & 0.6	 & 2.4	 & 0.6	 & 2.62	 & N \\ 
N1-1-004	 & 2012-05-11	 & 14.1	 & 3.89	 & 8.04	 & 8.93	 & 37	 & 11.5	 & 0.4	 & 13.4	 & 0.3	 & 4.2	 & 0.3	 & 2.62	 & N \\ 
N1-2-016	 & 2012-05-11	 & 16.7	 & 4.86	 & 7.50	 & 8.94	 & 35	 & 5.6	 & 0.5	 & 8.6	 & 0.4	 & 2.2	 & 0.3	 & 2.70	 & N \\ 
N1-2-017	 & 2012-05-11	 & 15.5	 & 5.40	 & 7.14	 & 8.95	 & 23	 & 12.4	 & 0.6	 & 13.0	 & 0.4	 & 1.9	 & 0.5	 & 2.82	 & N \\ 
N1-2-001	 & 2012-05-11	 & 13.2	 & 5.92	 & 6.93	 & 9.11	 & \textgreater40	 & 13.0	 & 0.3	 & 14.5	 & 0.2	 & 3.7	 & 0.2	 & 2.80	 & N \\ 
N1-2-021	 & 2012-05-11	 & 15.6	 & 4.57	 & 7.90	 & 9.12	 & 21	 & 8.7	 & 0.8	 & 12.2	 & 0.6	 & 3.6	 & 0.6	 & 2.63	 & N \\ 
N1-2-008	 & 2012-05-11	 & 15.1	 & 4.36	 & 8.06	 & 9.17	 & 30	 & 10.3	 & 0.5	 & 15.3	 & 0.3	 & 3.9	 & 0.4	 & 2.61	 & N \\ 
N1-1-009	 & 2012-05-11	 & 14.5	 & 2.10	 & 8.98	 & 9.22	 & 31	 & 11.9	 & 0.4	 & 15.2	 & 0.4	 & 3.8	 & 0.3	 & 2.58	 & N \\ 
N1-2-022	 & 2012-05-11	 & 15.9	 & 6.57	 & 6.49	 & 9.24	 & 17	 & 9.7	 & 1.0	 & 11.0	 & 0.6	 & 2.4	 & 0.7	 & 2.69	 & N \\ 
N1-2-031	 & 2012-05-11	 & 16.0	 & 5.55	 & 7.43	 & 9.28	 & 26	 & 7.4	 & 0.6	 & 11.0	 & 0.4	 & 3.5	 & 0.5	 & 2.75	 & N \\ 
N1-2-025	 & 2012-05-11	 & 15.9	 & 5.99	 & 7.12	 & 9.31	 & 7	 & 14.5	 & 2.4	 & 14.8	 & 2.0	 & 5.5	 & 1.4	 & 2.81	 & Y \\ 
N1-1-032	 & 2012-05-11	 & 15.7	 & 3.61	 & 8.58	 & 9.31	 & 22	 & 10.4	 & 0.7	 & 7.9	 & 0.6	 & 0.1	 & 0.5	 & 2.64	 & N \\ 
N1-1-029	 & 2012-05-11	 & 15.7	 & 3.20	 & 8.78	 & 9.34	 & 23	 & 7.9	 & 0.7	 & 10.3	 & 0.5	 & 1.3	 & 0.5	 & 2.63	 & N \\ 
N1-2-033	 & 2012-05-11	 & 16.1	 & 4.94	 & 7.95	 & 9.36	 & 17	 & 6.7	 & 0.9	 & 11.0	 & 0.7	 & 2.3	 & 0.6	 & 2.68	 & N \\ 
N1-2-027	 & 2012-05-11	 & 15.7	 & 6.59	 & 6.69	 & 9.39	 & 19	 & 8.1	 & 0.8	 & 10.8	 & 0.6	 & 3.1	 & 0.6	 & 2.69	 & N \\ 
NE1-1-003	 & 2014-05-14	 & 11.6	 & 8.61	 & 3.76	 & 9.40	 & \textgreater40	 & 4.4	 & 0.1	 & 8.5	 & 0.1	 & 0.6	 & 0.1	 & 2.55	 & N \\ 
N1-1-015	 & 2012-05-11	 & 15.3	 & 3.06	 & 8.91	 & 9.42	 & 28	 & 11.5	 & 0.5	 & 14.5	 & 0.4	 & 4.6	 & 0.4	 & 2.63	 & N \\ 
N1-1-018	 & 2012-05-11	 & 15.2	 & 1.63	 & 9.30	 & 9.44	 & 20	 & 10.4	 & 0.8	 & 16.2	 & 0.5	 & 3.8	 & 0.5	 & 2.57	 & N \\ 
N1-1-042	 & 2012-05-11	 & 16.0	 & 4.03	 & 8.55	 & 9.45	 & 12	 & 3.3	 & 1.7	 & 8.1	 & 1.0	 & 3.3	 & 0.9	 & 2.63	 & N \\ 
N1-1-010	 & 2012-05-11	 & 15.3	 & 1.94	 & 9.31	 & 9.51	 & 24	 & 11.1	 & 0.7	 & 14.4	 & 0.4	 & 3.3	 & 0.4	 & 2.56	 & N \\ 
NE1-1-002	 & 2014-05-14	 & 10.8	 & 8.58	 & 4.10	 & 9.51	 & \textgreater40	 & 13.6	 & 0.3	 & 19.9	 & 0.2	 & 3.9	 & 0.2	 & 2.59	 & N \\ 
N1-1-043	 & 2012-05-11	 & 16.3	 & 3.87	 & 8.69	 & 9.51	 & 8	 & 7.4	 & 1.7	 & 8.8	 & 1.3	 & 1.7	 & 1.3	 & 2.63	 & N \\ 
N1-2-024	 & 2012-05-11	 & 15.9	 & 5.11	 & 8.11	 & 9.58	 & 16	 & 9.1	 & 0.9	 & 13.3	 & 0.6	 & 3.0	 & 0.6	 & 2.67	 & N \\ 
NE1-1-026	 & 2014-05-14	 & 15.8	 & 9.39	 & 1.95	 & 9.59	 & 17	 & 9.2	 & 0.9	 & 10.2	 & 0.6	 & 2.8	 & 0.7	 & 2.44	 & N \\ 
NE1-1-025	 & 2014-05-14	 & 15.6	 & 9.18	 & 2.84	 & 9.61	 & \textgreater40	 & 9.0	 & 0.3	 & 11.8	 & 0.2	 & 3.4	 & 0.2	 & 2.38	 & N \\ 
N1-1-044	 & 2012-05-11	 & 15.8	 & 3.77	 & 8.91	 & 9.68	 & 27	 & 12.5	 & 0.6	 & 11.5	 & 0.5	 & 2.0	 & 0.4	 & 2.63	 & N \\ 
N1-1-024	 & 2012-05-11	 & 15.6	 & 3.38	 & 9.11	 & 9.71	 & 19	 & 7.0	 & 0.8	 & 9.4	 & 0.6	 & 4.0	 & 0.6	 & 2.63	 & N \\ 
N1-1-007	 & 2012-05-11	 & 14.4	 & 3.97	 & 8.89	 & 9.74	 & 36	 & 14.4	 & 0.4	 & 14.7	 & 0.3	 & 3.0	 & 0.3	 & 2.62	 & N \\ 
N1-2-011	 & 2012-05-11	 & 15.2	 & 6.65	 & 7.19	 & 9.79	 & 32	 & 10.0	 & 0.4	 & 12.7	 & 0.3	 & 3.3	 & 0.3	 & 2.82	 & N \\ 
N1-2-039	 & 2012-05-11	 & 16.6	 & 6.10	 & 7.69	 & 9.82	 & 11	 & 8.5	 & 1.5	 & 7.7	 & 0.9	 & 2.8	 & 1.0	 & 2.84	 & Y \\ 
NE1-1-032	 & 2014-05-14	 & 16.2	 & 9.61	 & 2.07	 & 9.83	 & 25	 & 7.4	 & 0.7	 & 10.2	 & 0.5	 & 3.4	 & 0.5	 & 2.42	 & N \\ 
NE1-1-011	 & 2014-05-14	 & 14.5	 & 9.36	 & 3.15	 & 9.87	 & \textgreater40	 & 12.1	 & 0.3	 & 14.4	 & 0.2	 & 3.9	 & 0.3	 & 2.37	 & N \\ 
N1-2-012	 & 2012-05-11	 & 15.4	 & 5.76	 & 8.04	 & 9.89	 & 30	 & 11.6	 & 0.6	 & 14.3	 & 0.4	 & 3.5	 & 0.4	 & 2.72	 & N \\ 
N1-1-035	 & 2012-05-11	 & 16.8	 & 3.68	 & 9.19	 & 9.90	 & 6	 & 11.9	 & 2.1	 & 16.9	 & 1.5	 & 7.7	 & 1.8	 & 2.63	 & N \\ 
NE1-1-009	 & 2014-05-14	 & 13.6	 & 8.98	 & 4.25	 & 9.93	 & \textgreater40	 & 13.2	 & 0.3	 & 17.7	 & 0.2	 & 4.7	 & 0.2	 & 2.57	 & N \\ 
N1-2-004	 & 2012-05-11	 & 13.8	 & 6.33	 & 7.68	 & 9.95	 & \textgreater40	 & 12.2	 & 0.3	 & 13.4	 & 0.3	 & 2.9	 & 0.2	 & 2.82	 & N \\ 
N1-1-003	 & 2012-05-11	 & 14.0	 & 4.25	 & 9.00	 & 9.95	 & \textgreater40	 & 10.5	 & 0.4	 & 13.1	 & 0.2	 & 3.2	 & 0.2	 & 2.64	 & N \\ 
N1-1-040	 & 2012-05-11	 & 15.4	 & 1.91	 & 9.78	 & 9.97	 & 7	 & 12.3	 & 1.8	 & 15.4	 & 1.4	 & 3.4	 & 1.5	 & 2.51	 & N \\ 
N1-1-013	 & 2012-05-11	 & 14.6	 & 3.92	 & 9.21	 & 10.01	 & 27	 & 12.7	 & 0.5	 & 15.4	 & 0.4	 & 4.6	 & 0.4	 & 2.63	 & N \\ 
NE1-1-038	 & 2014-05-14	 & 16.4	 & 9.66	 & 2.65	 & 10.01	 & 11	 & 8.9	 & 1.4	 & 11.6	 & 1.2	 & 3.4	 & 1.1	 & 2.35	 & N \\ 
N1-1-030	 & 2012-05-11	 & 16.0	 & 2.59	 & 9.69	 & 10.03	 & 13	 & 8.8	 & 1.4	 & 11.3	 & 0.9	 & 2.0	 & 0.8	 & 2.49	 & N \\ 
N1-1-026	 & 2012-05-11	 & 15.6	 & 3.23	 & 9.52	 & 10.05	 & 17	 & 10.1	 & 0.8	 & 11.7	 & 0.6	 & 1.9	 & 0.7	 & 2.57	 & N \\ 
NE1-1-007	 & 2014-05-14	 & 13.9	 & 9.68	 & 2.99	 & 10.13	 & 39	 & 13.5	 & 0.4	 & 15.8	 & 0.3	 & 4.5	 & 0.3	 & 2.35	 & N \\ 
NE1-1-043	 & 2014-05-14	 & 16.4	 & 9.38	 & 3.86	 & 10.15	 & 14	 & 9.2	 & 1.3	 & 12.7	 & 0.7	 & 3.8	 & 0.8	 & 2.41	 & N \\ 
N1-1-025	 & 2012-05-11	 & 15.7	 & 3.06	 & 9.69	 & 10.17	 & 12	 & 8.5	 & 1.2	 & 10.3	 & 0.8	 & 4.9	 & 1.0	 & 2.51	 & N \\ 
NE1-1-037	 & 2014-05-14	 & 16.3	 & 9.51	 & 3.63	 & 10.17	 & 21	 & 6.5	 & 0.7	 & 8.3	 & 0.4	 & 1.9	 & 0.5	 & 2.35	 & N \\ 
N1-1-041	 & 2012-05-11	 & 14.4	 & 2.11	 & 9.96	 & 10.18	 & 32	 & 13.4	 & 0.4	 & 17.6	 & 0.3	 & 3.5	 & 0.3	 & 2.49	 & N \\ 
N1-2-003	 & 2012-05-11	 & 13.5	 & 5.48	 & 8.60	 & 10.20	 & \textgreater40	 & 14.4	 & 0.4	 & 16.8	 & 0.3	 & 4.9	 & 0.2	 & 2.66	 & N \\ 
N1-2-006	 & 2012-05-11	 & 14.9	 & 6.08	 & 8.20	 & 10.20	 & 36	 & 10.2	 & 0.4	 & 12.1	 & 0.3	 & 3.9	 & 0.3	 & 2.68	 & N \\ 
N1-2-018	 & 2012-05-11	 & 15.5	 & 5.02	 & 8.90	 & 10.22	 & 16	 & 10.5	 & 0.8	 & 14.6	 & 0.7	 & 2.9	 & 0.7	 & 2.66	 & N \\ 
N1-1-005	 & 2012-05-11	 & 13.7	 & 3.50	 & 9.63	 & 10.25	 & \textgreater40	 & 11.0	 & 0.4	 & 14.4	 & 0.2	 & 4.3	 & 0.3	 & 2.58	 & N \\ 
NE1-1-019	 & 2014-05-14	 & 16.8	 & 9.98	 & 2.39	 & 10.26	 & 30	 & 9.8	 & 0.5	 & 12.0	 & 0.3	 & 3.0	 & 0.3	 & 2.35	 & N \\ 
N1-2-002	 & 2012-05-11	 & 13.3	 & 6.67	 & 7.99	 & 10.41	 & \textgreater40	 & 14.8	 & 0.3	 & 17.4	 & 0.2	 & 4.5	 & 0.2	 & 2.75	 & N \\ 
N1-1-008	 & 2012-05-11	 & 14.4	 & 3.00	 & 10.00	 & 10.44	 & 30	 & 11.8	 & 0.5	 & 13.2	 & 0.3	 & 2.7	 & 0.4	 & 2.50	 & N \\ 
N1-1-021	 & 2012-05-11	 & 15.4	 & 2.78	 & 10.15	 & 10.53	 & 12	 & 10.9	 & 1.4	 & 14.3	 & 0.9	 & 1.9	 & 0.9	 & 2.49	 & N \\ 
NE1-1-034	 & 2014-05-14	 & 16.0	 & 9.84	 & 4.02	 & 10.63	 & 25	 & 7.9	 & 0.7	 & 10.7	 & 0.4	 & 3.5	 & 0.4	 & 2.37	 & N \\ 
NE1-1-015	 & 2014-05-14	 & 15.5	 & 9.64	 & 4.65	 & 10.64	 & 27	 & 4.0	 & 0.6	 & 8.7	 & 0.5	 & 0.4	 & 0.5	 & 2.46	 & N \\ 
NE1-1-016	 & 2014-05-14	 & 15.3	 & 10.18	 & 3.10	 & 10.64	 & 33	 & 9.8	 & 0.5	 & 12.1	 & 0.3	 & 4.2	 & 0.3	 & 2.34	 & N \\ 
NE1-1-005	 & 2014-05-14	 & 12.6	 & 9.43	 & 4.95	 & 10.65	 & \textgreater40	 & 12.2	 & 0.3	 & 18.0	 & 0.2	 & 4.0	 & 0.2	 & 2.48	 & N \\ 
NE1-1-024	 & 2014-05-14	 & 15.5	 & 10.38	 & 2.68	 & 10.72	 & 31	 & 11.6	 & 0.5	 & 14.3	 & 0.3	 & 4.0	 & 0.3	 & 2.34	 & N \\ 
NE1-1-041	 & 2014-05-14	 & 16.5	 & 10.09	 & 3.64	 & 10.73	 & 15	 & 8.8	 & 0.9	 & 11.6	 & 0.7	 & 3.2	 & 0.7	 & 2.35	 & N \\ 
N2-1-008	 & 2012-05-13	 & 13.8	 & 3.34	 & 10.27	 & 10.80	 & 18	 & 15.0	 & 0.8	 & 15.2	 & 0.6	 & 0.7	 & 0.7	 & 2.54	 & N \\ 
NE1-1-021	 & 2014-05-14	 & 15.3	 & 9.84	 & 4.49	 & 10.82	 & 28	 & 5.3	 & 0.7	 & 6.9	 & 0.5	 & 1.9	 & 0.4	 & 2.44	 & Y \\ 
NE1-1-044	 & 2014-05-14	 & 17.3	 & 10.51	 & 2.62	 & 10.83	 & 18	 & 11.9	 & 0.7	 & 13.3	 & 0.6	 & 4.1	 & 0.6	 & 2.35	 & Y \\ 
NE1-1-030	 & 2014-05-14	 & 15.7	 & 10.09	 & 4.26	 & 10.95	 & 25	 & 9.2	 & 0.6	 & 11.6	 & 0.5	 & 2.9	 & 0.5	 & 2.43	 & N \\ 
N2-2-002	 & 2012-05-13	 & 13.1	 & 6.04	 & 9.14	 & 10.96	 & 18	 & 14.8	 & 0.7	 & 16.8	 & 0.5	 & 3.3	 & 0.7	 & 2.65	 & N \\ 
NE1-1-018	 & 2014-05-14	 & 15.4	 & 10.37	 & 3.57	 & 10.96	 & 39	 & 7.5	 & 0.5	 & 10.7	 & 0.3	 & 3.0	 & 0.3	 & 2.37	 & N \\ 
N2-2-008	 & 2012-05-13	 & 14.1	 & 6.59	 & 8.76	 & 10.96	 & 10	 & 11.8	 & 1.5	 & 15.8	 & 1.1	 & 2.9	 & 1.2	 & 2.68	 & N \\ 
NE1-1-022	 & 2014-05-14	 & 15.5	 & 10.60	 & 3.06	 & 11.04	 & 31	 & 5.8	 & 0.6	 & 9.1	 & 0.3	 & 1.1	 & 0.4	 & 2.35	 & N \\ 
N2-2-006	 & 2012-05-13	 & 14.4	 & 7.25	 & 8.52	 & 11.19	 & 13	 & 10.6	 & 1.0	 & 16.2	 & 0.8	 & 5.1	 & 0.9	 & 2.79	 & N \\ 
N2-1-015	 & 2012-05-13	 & 15.8	 & 3.08	 & 10.83	 & 11.26	 & 9	 & 10.2	 & 1.3	 & 15.8	 & 1.2	 & 6.8	 & 1.2	 & 2.51	 & N \\ 
N2-1-004	 & 2012-05-13	 & 13.2	 & 3.63	 & 10.68	 & 11.28	 & 37	 & 11.8	 & 0.5	 & 14.1	 & 0.3	 & 3.8	 & 0.3	 & 2.60	 & N \\ 
NE1-1-040	 & 2014-05-14	 & 16.5	 & 10.44	 & 4.37	 & 11.32	 & 21	 & 8.4	 & 0.8	 & 10.3	 & 0.6	 & 3.2	 & 0.5	 & 2.43	 & N \\ 
NE1-1-010	 & 2014-05-14	 & 14.2	 & 10.80	 & 3.67	 & 11.41	 & \textgreater40	 & 9.3	 & 0.4	 & 12.5	 & 0.3	 & 3.2	 & 0.3	 & 2.45	 & N \\ 
N2-1-006	 & 2012-05-13	 & 13.6	 & 4.98	 & 10.27	 & 11.42	 & 27	 & 14.4	 & 0.5	 & 15.4	 & 0.4	 & 5.1	 & 0.4	 & 2.70	 & N \\ 
NE1-1-031	 & 2014-05-14	 & 15.6	 & 10.23	 & 5.12	 & 11.44	 & 30	 & 9.6	 & 0.5	 & 12.5	 & 0.3	 & 3.6	 & 0.4	 & 2.43	 & N \\ 
NE1-1-028	 & 2014-05-14	 & 15.7	 & 10.76	 & 4.24	 & 11.56	 & 31	 & 7.1	 & 0.6	 & 11.5	 & 0.3	 & 2.9	 & 0.4	 & 2.43	 & N \\ 
NE1-1-008	 & 2014-05-14	 & 13.7	 & 10.54	 & 4.92	 & 11.63	 & \textgreater40	 & 11.6	 & 0.4	 & 16.4	 & 0.2	 & 3.7	 & 0.2	 & 2.44	 & N \\ 
NE1-1-029	 & 2014-05-14	 & 15.8	 & 10.84	 & 4.54	 & 11.75	 & 29	 & 1.6	 & 0.7	 & 5.7	 & 0.4	 & -0.3	 & 0.5	 & 2.41	 & N \\ 
N2-1-002	 & 2012-05-13	 & 12.2	 & 4.40	 & 11.00	 & 11.84	 & \textgreater40	 & 6.5	 & 0.2	 & 6.2	 & 0.1	 & 0.4	 & 0.1	 & 2.66	 & N \\ 
NE1-1-014	 & 2014-05-14	 & 15.5	 & 11.51	 & 2.86	 & 11.86	 & 38	 & 10.4	 & 0.5	 & 12.6	 & 0.3	 & 3.4	 & 0.3	 & 2.43	 & N \\ 
N2-1-009	 & 2012-05-13	 & 13.9	 & 5.00	 & 10.77	 & 11.87	 & 19	 & 13.1	 & 0.8	 & 15.0	 & 0.5	 & 3.5	 & 0.6	 & 2.75	 & N \\ 
NE1-1-033	 & 2014-05-14	 & 15.7	 & 10.55	 & 5.54	 & 11.91	 & 20	 & 10.9	 & 0.8	 & 15.2	 & 0.6	 & 4.1	 & 0.5	 & 2.42	 & N \\ 
N2-2-004	 & 2012-05-13	 & 16.9	 & 6.83	 & 9.80	 & 11.95	 & 15	 & 17.4	 & 0.9	 & 20.5	 & 0.6	 & 6.3	 & 0.7	 & 2.68	 & N \\ 
NE1-1-012	 & 2014-05-14	 & 14.8	 & 11.11	 & 4.41	 & 11.95	 & \textgreater40	 & 7.7	 & 0.4	 & 11.8	 & 0.3	 & 2.6	 & 0.2	 & 2.36	 & N \\ 
NE1-1-017	 & 2014-05-14	 & 15.2	 & 10.80	 & 5.18	 & 11.97	 & 25	 & 11.0	 & 0.6	 & 15.6	 & 0.4	 & 4.2	 & 0.5	 & 2.41	 & N \\ 
N2-1-007	 & 2012-05-13	 & 14.3	 & 3.36	 & 11.59	 & 12.06	 & 24	 & 9.7	 & 0.7	 & 13.1	 & 0.4	 & 2.5	 & 0.4	 & 2.60	 & N \\ 
NE1-1-013	 & 2014-05-14	 & 15.1	 & 11.08	 & 4.79	 & 12.07	 & \textgreater40	 & 6.1	 & 0.4	 & 11.6	 & 0.3	 & 2.5	 & 0.3	 & 2.40	 & N \\ 
N2-1-003	 & 2012-05-13	 & 12.9	 & 3.59	 & 11.53	 & 12.08	 & \textgreater40	 & 7.2	 & 0.4	 & 9.8	 & 0.3	 & 0.8	 & 0.3	 & 2.61	 & N \\ 
NE1-1-001	 & 2014-05-14	 & 10.7	 & 11.54	 & 3.73	 & 12.13	 & \textgreater40	 & 16.2	 & 0.3	 & 20.8	 & 0.2	 & 5.0	 & 0.2	 & 2.42	 & N \\ 
NE1-1-020	 & 2014-05-14	 & 15.2	 & 11.32	 & 4.38	 & 12.14	 & 19	 & 9.0	 & 0.9	 & 11.3	 & 0.7	 & 5.6	 & 0.6	 & 2.36	 & N \\ 
N2-2-001	 & 2012-05-13	 & 11.5	 & 7.03	 & 9.99	 & 12.21	 & 29	 & 16.2	 & 0.5	 & 19.8	 & 0.4	 & 6.2	 & 0.4	 & 2.72	 & N \\ 
N2-1-017	 & 2012-05-13	 & 15.7	 & 4.69	 & 11.67	 & 12.58	 & 5	 & \nodata	 & \nodata	 & \nodata	 & \nodata	 & \nodata	 & \nodata	 & 2.73	 & N \\ 
N2-1-016	 & 2012-05-13	 & 15.2	 & 4.25	 & 11.88	 & 12.62	 & 9	 & 7.8	 & 1.7	 & 13.6	 & 1.0	 & 5.3	 & 1.2	 & 2.63	 & N \\ 
N2-1-010	 & 2012-05-13	 & 13.8	 & 5.83	 & 11.30	 & 12.71	 & 3	 & \nodata	 & \nodata	 & \nodata	 & \nodata	 & \nodata	 & \nodata	 & 2.96	 & N \\ 
N2-1-018	 & 2012-05-13	 & 15.8	 & 4.70	 & 11.89	 & 12.79	 & 5	 & \nodata	 & \nodata	 & \nodata	 & \nodata	 & \nodata	 & \nodata	 & 2.71	 & N \\ 
N2-1-005	 & 2012-05-13	 & 13.5	 & 5.46	 & 11.60	 & 12.82	 & 17	 & 11.8	 & 0.8	 & 15.2	 & 0.6	 & 4.2	 & 0.7	 & 2.92	 & N \\ 
N2-1-014	 & 2012-05-13	 & 15.4	 & 3.36	 & 12.39	 & 12.84	 & 11	 & 8.2	 & 1.4	 & 11.2	 & 0.9	 & 6.4	 & 0.9	 & 2.55	 & N \\ 
N2-1-001	 & 2012-05-13	 & 11.8	 & 5.79	 & 11.57	 & 12.94	 & \textgreater40	 & 16.2	 & 0.3	 & 18.7	 & 0.2	 & 5.3	 & 0.2	 & 2.96	 & N \\ 
N2-2-003	 & 2012-05-13	 & 13.5	 & 7.04	 & 10.91	 & 12.98	 & 12	 & 12.6	 & 1.2	 & 15.1	 & 0.8	 & 4.5	 & 0.8	 & 2.73	 & N \\ 
N2-1-012	 & 2012-05-13	 & 14.4	 & 4.52	 & 12.47	 & 13.26	 & 23	 & 11.8	 & 0.6	 & 14.6	 & 0.4	 & 3.9	 & 0.4	 & 2.63	 & N \\ 
N2-2-005	 & 2012-05-13	 & 13.7	 & 7.80	 & 10.78	 & 13.31	 & 11	 & 15.0	 & 1.3	 & 29.7	 & 0.7	 & 2.1	 & 1.0	 & 2.78	 & N \\ 
E5-2-016	 & 2013-05-23	 & 14.9	 & 12.93	 & -3.34	 & 13.35	 & 16	 & 12.8	 & 0.9	 & 16.5	 & 0.5	 & 5.5	 & 0.6	 & 2.81	 & N \\ 
E5-2-025	 & 2013-05-23	 & 15.8	 & 13.46	 & -1.70	 & 13.57	 & 16	 & 7.9	 & 1.0	 & 12.9	 & 0.6	 & 4.6	 & 0.6	 & 2.74	 & N \\ 
N2-1-011	 & 2012-05-13	 & 15.2	 & 3.68	 & 13.14	 & 13.64	 & 10	 & 9.4	 & 1.2	 & 11.8	 & 1.0	 & 3.7	 & 1.1	 & 2.58	 & N \\ 
E5-2-002	 & 2013-05-23	 & 13.5	 & 13.34	 & -3.02	 & 13.68	 & 25	 & 15.7	 & 0.6	 & 17.5	 & 0.4	 & 4.5	 & 0.4	 & 2.83	 & N \\ 
E5-2-006	 & 2013-05-23	 & 14.2	 & 13.29	 & -3.38	 & 13.72	 & 36	 & 12.0	 & 0.4	 & 13.9	 & 0.3	 & 2.6	 & 0.3	 & 2.81	 & N \\ 
E5-1-009	 & 2012-05-07	 & 14.8	 & 13.83	 & -0.29	 & 13.84	 & 39	 & 7.0	 & 0.4	 & 13.5	 & 0.3	 & 3.3	 & 0.3	 & 2.70	 & N \\ 
E5-2-026	 & 2013-05-23	 & 15.9	 & 13.78	 & -1.33	 & 13.85	 & 22	 & 8.5	 & 0.8	 & 14.5	 & 0.4	 & 2.9	 & 0.5	 & 2.76	 & N \\ 
E5-2-028	 & 2013-05-23	 & 16.2	 & 13.81	 & -1.69	 & 13.91	 & 25	 & 6.0	 & 0.6	 & 10.3	 & 0.4	 & 3.7	 & 0.4	 & 2.78	 & N \\ 
E5-2-022	 & 2013-05-23	 & 16.0	 & 13.62	 & -3.17	 & 13.99	 & 10	 & 10.6	 & 1.5	 & 13.4	 & 1.0	 & 4.0	 & 1.0	 & 2.75	 & N \\ 
E5-2-003	 & 2013-05-23	 & 13.9	 & 13.76	 & -2.86	 & 14.05	 & \textgreater40	 & 5.0	 & 0.4	 & 9.6	 & 0.2	 & 2.2	 & 0.2	 & 2.77	 & N \\ 
E5-1-012	 & 2012-05-07	 & 15.3	 & 14.12	 & -0.04	 & 14.12	 & 36	 & 12.4	 & 0.4	 & 16.8	 & 0.3	 & 3.7	 & 0.3	 & 2.59	 & N \\ 
E5-1-017	 & 2012-05-07	 & 15.9	 & 14.25	 & 0.04	 & 14.25	 & 26	 & 8.5	 & 0.6	 & 11.2	 & 0.4	 & 2.7	 & 0.5	 & 2.59	 & Y \\ 
E5-1-021	 & 2012-05-07	 & 15.5	 & 14.30	 & -0.47	 & 14.31	 & 26	 & 7.5	 & 0.6	 & 11.3	 & 0.4	 & 2.4	 & 0.5	 & 2.68	 & N \\ 
E5-2-017	 & 2013-05-23	 & 16.3	 & 13.95	 & -3.27	 & 14.32	 & 28	 & 2.0	 & 0.7	 & 3.0	 & 0.4	 & 1.3	 & 0.4	 & 2.77	 & N \\ 
E5-2-032	 & 2013-05-23	 & 17.0	 & 13.90	 & -3.46	 & 14.32	 & 12	 & 8.3	 & 1.2	 & 7.6	 & 1.0	 & 4.3	 & 1.0	 & 2.79	 & N \\ 
E5-2-005	 & 2013-05-23	 & 14.1	 & 14.19	 & -2.05	 & 14.33	 & 38	 & 10.0	 & 0.4	 & 12.5	 & 0.3	 & 3.9	 & 0.3	 & 2.83	 & N \\ 
E5-2-027	 & 2013-05-23	 & 15.7	 & 14.34	 & -1.27	 & 14.40	 & 21	 & 7.2	 & 0.7	 & 11.5	 & 0.6	 & 3.2	 & 0.5	 & 2.83	 & N \\ 
E5-2-015	 & 2013-05-23	 & 15.6	 & 14.25	 & -2.72	 & 14.51	 & 24	 & 9.5	 & 0.6	 & 11.3	 & 0.4	 & 3.6	 & 0.4	 & 2.80	 & N \\ 
E5-2-013	 & 2013-05-23	 & 15.1	 & 14.40	 & -1.79	 & 14.51	 & 23	 & 10.2	 & 0.7	 & 12.6	 & 0.4	 & 3.4	 & 0.5	 & 2.82	 & N \\ 
E5-1-030	 & 2012-05-07	 & 16.6	 & 14.52	 & -0.13	 & 14.52	 & 17	 & 9.7	 & 0.9	 & 13.1	 & 0.6	 & 5.6	 & 0.6	 & 2.59	 & N \\ 
E5-1-019	 & 2012-05-07	 & 15.6	 & 14.51	 & 0.91	 & 14.54	 & 32	 & 9.8	 & 0.5	 & 13.1	 & 0.4	 & 4.1	 & 0.3	 & 2.54	 & N \\ 
E5-1-025	 & 2012-05-07	 & 16.2	 & 14.59	 & 0.45	 & 14.60	 & 19	 & 9.1	 & 0.7	 & 9.1	 & 0.6	 & 2.8	 & 0.6	 & 2.56	 & N \\ 
E5-1-002	 & 2012-05-07	 & 12.8	 & 14.52	 & 1.47	 & 14.60	 & \textgreater40	 & 11.0	 & 0.3	 & 17.2	 & 0.2	 & 4.1	 & 0.2	 & 2.47	 & N \\ 
E5-2-007	 & 2013-05-23	 & 14.2	 & 14.43	 & -2.53	 & 14.65	 & 34	 & 11.7	 & 0.4	 & 14.4	 & 0.3	 & 3.8	 & 0.3	 & 2.81	 & N \\ 
E5-2-009	 & 2013-05-23	 & 14.8	 & 14.49	 & -2.26	 & 14.67	 & 38	 & 10.4	 & 0.4	 & 14.7	 & 0.3	 & 3.9	 & 0.3	 & 2.83	 & N \\ 
E5-1-015	 & 2012-05-07	 & 15.7	 & 14.78	 & -0.36	 & 14.79	 & 39	 & 7.3	 & 0.4	 & 10.1	 & 0.3	 & 2.7	 & 0.3	 & 2.68	 & N \\ 
E5-1-004	 & 2012-05-07	 & 14.0	 & 14.79	 & 0.84	 & 14.82	 & 32	 & 12.8	 & 0.4	 & 17.5	 & 0.3	 & 5.8	 & 0.3	 & 2.53	 & N \\ 
E5-2-037	 & 2013-05-23	 & 17.9\tablenotemark{b}	 & 14.34	 & -3.75	 & 14.83	 & 3	 & \nodata	 & \nodata	 & \nodata	 & \nodata	 & \nodata	 & \nodata	 & 2.70	 & N \\ 
E5-2-029	 & 2013-05-23	 & 16.2	 & 14.78	 & -1.92	 & 14.90	 & 21	 & 8.0	 & 0.7	 & 11.3	 & 0.5	 & 3.2	 & 0.5	 & 2.79	 & N \\ 
E5-2-020	 & 2013-05-23	 & 15.7	 & 14.65	 & -2.91	 & 14.94	 & 36	 & 8.1	 & 0.4	 & 11.6	 & 0.3	 & 4.0	 & 0.3	 & 2.72	 & N \\ 
E5-1-022	 & 2012-05-07	 & 15.8	 & 15.01	 & 0.08	 & 15.01	 & 28	 & 9.7	 & 0.5	 & 13.5	 & 0.4	 & 2.8	 & 0.4	 & 2.58	 & N \\ 
E5-1-003	 & 2012-05-07	 & 13.4	 & 15.00	 & 1.06	 & 15.03	 & \textgreater40	 & 11.7	 & 0.4	 & 16.0	 & 0.3	 & 5.2	 & 0.3	 & 2.50	 & N \\ 
E5-1-008	 & 2012-05-07	 & 15.2	 & 15.02	 & 0.68	 & 15.04	 & 36	 & 8.1	 & 0.4	 & 11.3	 & 0.3	 & 3.1	 & 0.3	 & 2.54	 & N \\ 
E5-2-039	 & 2013-05-23	 & 16.6	 & 14.99	 & -2.24	 & 15.16	 & 15	 & 10.0	 & 1.0	 & 8.9	 & 0.7	 & 2.9	 & 0.8	 & 2.74	 & N \\ 
E5-1-001	 & 2012-05-07	 & 12.1	 & 15.17	 & -0.49	 & 15.18	 & \textgreater40	 & 14.5	 & 0.4	 & 18.3	 & 0.2	 & 6.2	 & 0.2	 & 2.66	 & N \\ 
E5-1-007	 & 2012-05-07	 & 15.5	 & 15.25	 & -0.23	 & 15.25	 & \textgreater40	 & 10.4	 & 0.4	 & 14.1	 & 0.3	 & 4.4	 & 0.3	 & 2.61	 & Y \\ 
E5-2-008	 & 2013-05-23	 & 14.6\tablenotemark{b}	 & 14.95	 & -3.23	 & 15.30	 & 34	 & 10.1	 & 0.4	 & 11.3	 & 0.3	 & 3.5	 & 0.3	 & 2.69	 & N \\ 
E5-1-042	 & 2012-05-07	 & 15.2	 & 15.28	 & 1.05	 & 15.31	 & 35	 & 9.3	 & 0.5	 & 11.7	 & 0.4	 & 1.1	 & 0.4	 & 2.50	 & N \\ 
E5-1-010	 & 2012-05-07	 & 15.3	 & 15.30	 & 1.48	 & 15.37	 & 31	 & 9.7	 & 0.6	 & 13.7	 & 0.3	 & 3.7	 & 0.4	 & 2.50	 & N \\ 
E5-1-032	 & 2012-05-07	 & 16.9	 & 15.37	 & 0.60	 & 15.38	 & 17	 & 7.6	 & 0.9	 & 8.0	 & 0.6	 & 1.3	 & 0.7	 & 2.55	 & N \\ 
E5-2-004	 & 2013-05-23	 & 13.9	 & 15.01	 & -3.50	 & 15.41	 & \textgreater40	 & 11.3	 & 0.3	 & 14.1	 & 0.2	 & 4.0	 & 0.2	 & 2.68	 & N \\ 
E5-1-011	 & 2012-05-07	 & 15.5	 & 15.39	 & -0.81	 & 15.41	 & 33	 & 11.2	 & 0.5	 & 13.9	 & 0.4	 & 4.6	 & 0.3	 & 2.65	 & N \\ 
E5-2-018	 & 2013-05-23	 & 15.5	 & 15.24	 & -2.37	 & 15.42	 & 32	 & 10.1	 & 0.5	 & 12.0	 & 0.4	 & 4.1	 & 0.3	 & 2.69	 & N \\ 
E5-1-018	 & 2012-05-07	 & 15.5	 & 15.45	 & -0.31	 & 15.45	 & 23	 & 7.0	 & 0.7	 & 7.7	 & 0.6	 & 1.5	 & 0.5	 & 2.61	 & Y \\ 
E5-1-024	 & 2012-05-07	 & 17.0\tablenotemark{b}	 & 15.45	 & -0.57	 & 15.46	 & 8	 & 14.2	 & 1.8	 & 17.3	 & 1.4	 & 6.4	 & 1.3	 & 2.62	 & N \\ 
E5-2-019	 & 2013-05-23	 & 15.6	 & 15.10	 & -3.87	 & 15.59	 & 37	 & 10.0	 & 0.4	 & 13.5	 & 0.3	 & 3.7	 & 0.3	 & 2.66	 & N \\ 
E5-1-026	 & 2012-05-07	 & 16.3	 & 15.61	 & 0.30	 & 15.61	 & 36	 & 8.0	 & 0.5	 & 8.7	 & 0.3	 & 2.1	 & 0.3	 & 2.58	 & N \\ 
E5-2-021	 & 2013-05-23	 & 15.7	 & 15.35	 & -3.33	 & 15.71	 & 25	 & 9.1	 & 0.7	 & 10.7	 & 0.4	 & 2.7	 & 0.4	 & 2.67	 & N \\ 
E5-1-005	 & 2012-05-07	 & 14.7	 & 15.66	 & 1.50	 & 15.73	 & \textgreater40	 & 2.3	 & 0.4	 & 3.8	 & 0.2	 & 1.1	 & 0.2	 & 2.47	 & N \\ 
E5-2-014	 & 2013-05-23	 & 15.3	 & 15.65	 & -1.86	 & 15.76	 & 28	 & 11.5	 & 0.6	 & 12.9	 & 0.4	 & 5.0	 & 0.4	 & 2.70	 & N \\ 
E5-1-016	 & 2012-05-07	 & 15.6	 & 15.75	 & -0.69	 & 15.77	 & \textgreater40	 & 9.3	 & 0.4	 & 11.9	 & 0.3	 & 3.9	 & 0.3	 & 2.64	 & N \\ 
E5-2-010	 & 2013-05-23	 & 15.1	 & 15.67	 & -2.25	 & 15.83	 & \textgreater40	 & 7.5	 & 0.3	 & 9.7	 & 0.2	 & 2.4	 & 0.2	 & 2.70	 & N \\ 
E5-1-033	 & 2012-05-07	 & 17.4	 & 15.89	 & 0.49	 & 15.90	 & 12	 & 10.6	 & 1.3	 & 8.9	 & 0.9	 & 3.3	 & 1.0	 & 2.56	 & N \\ 
E5-2-012	 & 2013-05-23	 & 15.5	 & 15.95	 & -1.62	 & 16.03	 & 32	 & 7.2	 & 0.6	 & 9.0	 & 0.4	 & 0.7	 & 0.4	 & 2.61	 & N \\ 
E5-2-001	 & 2013-05-23	 & 11.4	 & 15.47	 & -4.21	 & 16.04	 & \textgreater40	 & 14.5	 & 0.3	 & 19.6	 & 0.2	 & 5.3	 & 0.2	 & 2.67	 & N \\ 
E5-1-029	 & 2012-05-07	 & 16.5	 & 16.03	 & -0.67	 & 16.04	 & 22	 & 6.7	 & 0.8	 & 9.1	 & 0.5	 & 3.6	 & 0.5	 & 2.61	 & N \\ 
E5-2-011	 & 2013-05-23	 & 15.4	 & 15.72	 & -3.19	 & 16.05	 & \textgreater40	 & 8.4	 & 0.4	 & 11.1	 & 0.3	 & 3.7	 & 0.3	 & 2.66	 & N \\ 
E5-1-028	 & 2012-05-07	 & 16.6	 & 16.02	 & 1.41	 & 16.08	 & 14	 & 7.9	 & 0.9	 & 8.3	 & 0.8	 & 3.8	 & 1.0	 & 2.51	 & N \\ 
E5-1-014	 & 2012-05-07	 & 15.6	 & 16.53	 & 1.32	 & 16.58	 & 38	 & 9.6	 & 0.4	 & 11.4	 & 0.3	 & 4.0	 & 0.3	 & 2.54	 & N \\ 
E6-2-002	 & 2012-05-05	 & 13.5	 & 16.21	 & -3.57	 & 16.60	 & \textgreater40	 & 12.9	 & 0.3	 & 17.8	 & 0.2	 & 4.5	 & 0.2	 & 2.68	 & N \\ 
E6-2-016	 & 2012-05-05	 & 15.7	 & 16.03	 & -4.48	 & 16.65	 & 17	 & 8.8	 & 1.0	 & 14.3	 & 0.7	 & 4.9	 & 0.7	 & 2.73	 & N \\ 
E5-1-023	 & 2012-05-07	 & 16.4	 & 16.60	 & 1.52	 & 16.67	 & 11	 & 9.4	 & 1.3	 & 10.8	 & 0.9	 & 1.6	 & 1.1	 & 2.55	 & N \\ 
E6-2-022	 & 2012-05-05	 & 15.7	 & 16.63	 & -1.85	 & 16.73	 & 14	 & 11.3	 & 1.0	 & 16.9	 & 0.8	 & 5.2	 & 0.7	 & 2.61	 & Y \\ 
E6-2-007	 & 2012-05-05	 & 15.0	 & 16.58	 & -2.52	 & 16.77	 & 33	 & 6.4	 & 0.5	 & 12.0	 & 0.3	 & 2.3	 & 0.3	 & 2.63	 & N \\ 
E6-2-039	 & 2012-05-05	 & 17.3	 & 16.56	 & -2.84	 & 16.80	 & 6	 & 12.6	 & 2.1	 & 14.6	 & 2.0	 & 3.4	 & 1.9	 & 2.68	 & N \\ 
E5-1-006	 & 2012-05-07	 & 14.8	 & 16.80	 & 0.36	 & 16.81	 & 40	 & 11.2	 & 0.4	 & 13.7	 & 0.2	 & 3.8	 & 0.3	 & 2.60	 & N \\ 
E6-1-016	 & 2013-05-30	 & 15.4	 & 16.83	 & -0.96	 & 16.86	 & 18	 & 14.7	 & 0.8	 & 20.1	 & 0.6	 & 4.3	 & 0.6	 & 2.61	 & Y \\ 
E6-2-018	 & 2012-05-05	 & 15.6	 & 16.77	 & -1.77	 & 16.86	 & 12	 & 6.7	 & 1.2	 & 11.8	 & 0.8	 & 3.3	 & 0.9	 & 2.61	 & N \\ 
E6-1-036	 & 2013-05-30	 & 15.5	 & 16.84	 & -1.13	 & 16.88	 & 23	 & 7.8	 & 0.6	 & 13.4	 & 0.4	 & 3.5	 & 0.5	 & 2.63	 & N \\ 
E6-1-015	 & 2013-05-30	 & 15.7	 & 16.87	 & -0.78	 & 16.89	 & 13	 & 3.6	 & 1.2	 & 8.0	 & 0.9	 & 2.1	 & 0.9	 & 2.61	 & N \\ 
E6-2-004	 & 2012-05-05	 & 14.0	 & 16.54	 & -3.67	 & 16.94	 & \textgreater40	 & 13.6	 & 0.4	 & 17.7	 & 0.2	 & 3.8	 & 0.2	 & 2.73	 & N \\ 
E6-2-015	 & 2012-05-05	 & 15.4	 & 16.77	 & -2.43	 & 16.94	 & 29	 & 10.4	 & 0.5	 & 15.5	 & 0.3	 & 4.1	 & 0.4	 & 2.67	 & N \\ 
E6-1-011	 & 2013-05-30	 & 15.0	 & 17.00	 & -0.96	 & 17.03	 & 31	 & 9.4	 & 0.5	 & 14.2	 & 0.3	 & 3.3	 & 0.3	 & 2.62	 & N \\ 
E6-1-032	 & 2013-05-30	 & 17.6	 & 17.29	 & 0.16	 & 17.29	 & 11	 & 8.1	 & 1.3	 & 11.3	 & 1.0	 & 2.5	 & 1.1	 & 2.64	 & N \\ 
E6-2-041	 & 2012-05-05	 & 16.8	 & 16.96	 & -3.79	 & 17.38	 & 11	 & 8.4	 & 1.4	 & 6.7	 & 1.0	 & -0.1	 & 1.2	 & 2.74	 & N \\ 
E6-2-032	 & 2012-05-05	 & 16.9	 & 17.30	 & -2.53	 & 17.48	 & 18	 & 8.2	 & 0.7	 & 11.6	 & 0.5	 & 3.5	 & 0.6	 & 2.73	 & Y \\ 
E6-2-006	 & 2012-05-05	 & 14.8	 & 17.24	 & -3.20	 & 17.54	 & \textgreater40	 & 11.9	 & 0.4	 & 12.4	 & 0.3	 & 2.3	 & 0.3	 & 2.74	 & N \\ 
E6-1-009	 & 2013-05-30	 & 14.9	 & 17.56	 & -0.47	 & 17.57	 & \textgreater40	 & 7.9	 & 0.3	 & 11.4	 & 0.2	 & 2.6	 & 0.2	 & 2.67	 & N \\ 
E6-1-030	 & 2013-05-30	 & 16.7	 & 17.57	 & 0.37	 & 17.57	 & 13	 & 8.8	 & 1.2	 & 11.2	 & 0.7	 & 2.5	 & 0.9	 & 2.65	 & N \\ 
E6-2-036	 & 2012-05-05	 & 16.9	 & 17.49	 & -2.11	 & 17.62	 & 9	 & 8.7	 & 1.8	 & 13.5	 & 1.2	 & 5.0	 & 1.1	 & 2.72	 & N \\ 
E6-2-014	 & 2012-05-05	 & 15.5	 & 17.45	 & -2.61	 & 17.65	 & 29	 & 12.6	 & 0.5	 & 14.4	 & 0.4	 & 4.3	 & 0.4	 & 2.73	 & N \\ 
E6-1-020	 & 2013-05-30	 & 15.5	 & 17.63	 & -1.19	 & 17.67	 & 20	 & 11.5	 & 0.7	 & 13.6	 & 0.5	 & 3.6	 & 0.5	 & 2.67	 & N \\ 
E6-1-028	 & 2013-05-30	 & 16.5	 & 17.71	 & -0.61	 & 17.72	 & 9	 & 6.7	 & 1.8	 & 12.2	 & 1.1	 & 3.9	 & 1.3	 & 2.67	 & N \\ 
E6-2-005	 & 2012-05-05	 & 14.7	 & 17.49	 & -2.98	 & 17.74	 & 36	 & 12.6	 & 0.4	 & 15.4	 & 0.3	 & 4.5	 & 0.3	 & 2.74	 & N \\ 
E6-2-026	 & 2012-05-05	 & 15.8	 & 17.34	 & -3.80	 & 17.75	 & 15	 & 11.0	 & 1.0	 & 13.9	 & 0.7	 & 3.3	 & 0.6	 & 2.74	 & N \\ 
E6-2-037	 & 2012-05-05	 & 16.9	 & 17.53	 & -3.54	 & 17.88	 & 13	 & 6.7	 & 1.2	 & 8.5	 & 0.9	 & 2.6	 & 0.9	 & 2.74	 & N \\ 
E6-1-001	 & 2013-05-30	 & 12.0	 & 17.88	 & 0.89	 & 17.90	 & \textgreater40	 & 15.1	 & 0.3	 & 18.1	 & 0.2	 & 4.2	 & 0.2	 & 2.66	 & N \\ 
E6-1-005	 & 2013-05-30	 & 13.8	 & 17.91	 & -1.02	 & 17.94	 & \textgreater40	 & 7.5	 & 0.3	 & 11.4	 & 0.2	 & 2.6	 & 0.2	 & 2.68	 & N \\ 
E6-1-029	 & 2013-05-30	 & 16.1\tablenotemark{b}	 & 17.97	 & -0.85	 & 17.99	 & 15	 & 6.2	 & 1.1	 & 11.2	 & 0.8	 & 2.8	 & 0.8	 & 2.68	 & Y \\ 
E6-1-002	 & 2013-05-30	 & 12.8	 & 17.95	 & -1.80	 & 18.04	 & \textgreater40	 & 11.8	 & 0.3	 & 16.0	 & 0.2	 & 4.2	 & 0.2	 & 2.68	 & N \\ 
E6-1-021	 & 2013-05-30	 & 15.6	 & 18.06	 & -1.42	 & 18.12	 & 23	 & 9.9	 & 0.7	 & 11.6	 & 0.4	 & 3.2	 & 0.5	 & 2.68	 & N \\ 
E6-2-012	 & 2012-05-05	 & 15.2	 & 17.87	 & -3.05	 & 18.13	 & 26	 & 11.3	 & 0.6	 & 12.2	 & 0.5	 & 2.2	 & 0.4	 & 2.75	 & N \\ 
E6-1-010	 & 2013-05-30	 & 15.5	 & 18.18	 & -0.47	 & 18.19	 & \textgreater40	 & 6.1	 & 0.3	 & 8.5	 & 0.2	 & 2.2	 & 0.2	 & 2.69	 & N \\ 
E6-2-029	 & 2012-05-05	 & 16.1	 & 17.72	 & -4.17	 & 18.21	 & 14	 & 8.9	 & 1.0	 & 10.2	 & 0.8	 & 1.1	 & 0.9	 & 2.74	 & N \\ 
E6-2-020	 & 2012-05-05	 & 15.5	 & 17.83	 & -3.72	 & 18.21	 & 27	 & 11.2	 & 0.5	 & 11.9	 & 0.4	 & 3.2	 & 0.5	 & 2.74	 & N \\ 
E6-2-008	 & 2012-05-05	 & 15.2	 & 18.12	 & -2.94	 & 18.36	 & \textgreater40	 & 8.4	 & 0.4	 & 10.5	 & 0.2	 & 3.7	 & 0.3	 & 2.75	 & N \\ 
E6-2-009	 & 2012-05-05	 & 15.0	 & 17.80	 & -4.55	 & 18.37	 & \textgreater40	 & 9.9	 & 0.4	 & 11.7	 & 0.3	 & 2.6	 & 0.3	 & 2.75	 & N \\ 
E6-1-014	 & 2013-05-30	 & 15.4	 & 18.32	 & -1.51	 & 18.38	 & 32	 & 6.7	 & 0.5	 & 9.0	 & 0.4	 & 2.9	 & 0.4	 & 2.67	 & N \\ 
E6-1-008	 & 2013-05-30	 & 14.8	 & 18.40	 & -1.13	 & 18.43	 & 38	 & 11.1	 & 0.4	 & 12.9	 & 0.3	 & 4.3	 & 0.3	 & 2.69	 & N \\ 
E6-1-018	 & 2013-05-30	 & 15.5	 & 18.46	 & 0.44	 & 18.47	 & 25	 & 9.5	 & 0.6	 & 12.7	 & 0.4	 & 3.9	 & 0.4	 & 2.74	 & N \\ 
E6-2-038	 & 2012-05-05	 & 17.5\tablenotemark{b}	 & 18.05	 & -3.97	 & 18.48	 & 7	 & 6.4	 & 2.0	 & 4.1	 & 1.9	 & 0.7	 & 1.8	 & 2.74	 & N \\ 
E6-1-023	 & 2013-05-30	 & 16.3	 & 18.53	 & -0.10	 & 18.53	 & 19	 & 8.7	 & 0.8	 & 12.1	 & 0.6	 & 3.0	 & 0.6	 & 2.67	 & N \\ 
E6-1-031	 & 2013-05-30	 & 17.3	 & 18.53	 & -0.85	 & 18.55	 & 9	 & 8.2	 & 1.4	 & 9.0	 & 1.2	 & 3.4	 & 1.4	 & 2.68	 & N \\ 
E6-1-022	 & 2013-05-30	 & 16.1	 & 18.48	 & -1.81	 & 18.57	 & 18	 & 9.4	 & 0.8	 & 11.3	 & 0.6	 & 3.0	 & 0.6	 & 2.69	 & N \\ 
E6-2-033	 & 2012-05-05	 & 16.7	 & 17.97	 & -4.75	 & 18.58	 & 4	 & \nodata	 & \nodata	 & \nodata	 & \nodata	 & \nodata	 & \nodata	 & 2.75	 & N \\ 
E6-2-010	 & 2012-05-05	 & 15.6	 & 18.52	 & -2.28	 & 18.66	 & 31	 & 10.5	 & 0.5	 & 10.5	 & 0.4	 & 3.2	 & 0.4	 & 2.71	 & N \\ 
E6-2-017	 & 2012-05-05	 & 15.7	 & 18.50	 & -2.71	 & 18.70	 & 25	 & 7.2	 & 0.5	 & 11.6	 & 0.4	 & 3.6	 & 0.4	 & 2.75	 & N \\ 
E6-2-035	 & 2012-05-05	 & 16.6	 & 18.56	 & -2.44	 & 18.72	 & 21	 & 10.9	 & 0.9	 & 7.7	 & 0.5	 & 1.0	 & 0.6	 & 2.70	 & Y \\ 
E6-1-019	 & 2013-05-30	 & 15.6	 & 18.76	 & -0.24	 & 18.76	 & 28	 & 8.5	 & 0.5	 & 11.0	 & 0.4	 & 2.8	 & 0.4	 & 2.66	 & N \\ 
E6-1-012	 & 2013-05-30	 & 15.4	 & 18.72	 & -1.68	 & 18.80	 & 26	 & 12.1	 & 0.6	 & 13.7	 & 0.4	 & 3.9	 & 0.3	 & 2.67	 & N \\ 
E6-2-027	 & 2012-05-05	 & 15.9	 & 18.43	 & -3.84	 & 18.83	 & 22	 & 12.0	 & 0.7	 & 12.7	 & 0.5	 & 5.2	 & 0.5	 & 2.84	 & N \\ 
E6-2-023	 & 2012-05-05	 & 15.7	 & 18.55	 & -3.27	 & 18.83	 & 23	 & 8.0	 & 0.7	 & 11.0	 & 0.4	 & 3.7	 & 0.6	 & 2.79	 & N \\ 
E6-1-026	 & 2013-05-30	 & 15.6	 & 18.82	 & -1.37	 & 18.87	 & 22	 & 15.2	 & 0.7	 & 15.3	 & 0.5	 & 3.0	 & 0.5	 & 2.64	 & Y \\ 
E6-2-025	 & 2012-05-05	 & 15.9	 & 18.32	 & -4.60	 & 18.89	 & 16	 & 7.5	 & 0.9	 & 8.8	 & 0.7	 & 1.6	 & 0.7	 & 2.73	 & N \\ 
E6-2-034	 & 2012-05-05	 & 16.8	 & 18.67	 & -2.87	 & 18.89	 & 3	 & \nodata	 & \nodata	 & \nodata	 & \nodata	 & \nodata	 & \nodata	 & 2.75	 & N \\ 
E6-1-004	 & 2013-05-30	 & 13.3	 & 18.92	 & 0.21	 & 18.92	 & \textgreater40	 & 6.1	 & 0.3	 & 7.9	 & 0.2	 & 2.1	 & 0.2	 & 2.71	 & N \\ 
E6-1-024	 & 2013-05-30	 & 16.2	 & 18.94	 & -1.18	 & 18.98	 & 17	 & 10.7	 & 0.9	 & 13.3	 & 0.7	 & 4.2	 & 0.6	 & 2.66	 & N \\ 
E6-1-003	 & 2013-05-30	 & 13.3	 & 18.92	 & -1.56	 & 18.98	 & \textgreater40	 & 14.7	 & 0.3	 & 17.5	 & 0.2	 & 5.0	 & 0.2	 & 2.65	 & N \\ 
E6-1-027	 & 2013-05-30	 & 16.3\tablenotemark{b}	 & 18.98	 & 0.42	 & 18.99	 & 18	 & 6.3	 & 1.2	 & 7.8	 & 0.7	 & 2.7	 & 0.6	 & 2.69	 & N \\ 
E6-2-031	 & 2012-05-05	 & 16.6	 & 18.63	 & -3.88	 & 19.03	 & 5	 & \nodata	 & \nodata	 & \nodata	 & \nodata	 & \nodata	 & \nodata	 & 2.74	 & N \\ 
E6-2-021	 & 2012-05-05	 & 15.6	 & 18.83	 & -3.23	 & 19.10	 & 27	 & 11.5	 & 0.6	 & 12.4	 & 0.4	 & 3.0	 & 0.4	 & 2.76	 & N \\ 
E6-2-024	 & 2012-05-05	 & 16.0	 & 18.64	 & -4.57	 & 19.20	 & 20	 & 11.4	 & 0.6	 & 11.8	 & 0.6	 & 1.5	 & 0.7	 & 2.70	 & N \\ 
E6-2-019	 & 2012-05-05	 & 15.9	 & 18.80	 & -3.97	 & 19.22	 & 25	 & 10.4	 & 0.6	 & 10.7	 & 0.5	 & 2.4	 & 0.5	 & 2.70	 & N \\ 
E6-2-003	 & 2012-05-05	 & 14.0	 & 19.07	 & -2.56	 & 19.24	 & \textgreater40	 & 14.3	 & 0.4	 & 16.9	 & 0.2	 & 3.9	 & 0.3	 & 2.69	 & N \\ 
E6-2-030	 & 2012-05-05	 & 16.3	 & 18.91	 & -3.72	 & 19.28	 & 12	 & 10.0	 & 1.4	 & 11.9	 & 0.9	 & 2.9	 & 0.8	 & 2.74	 & N \\ 
E6-2-001	 & 2012-05-05	 & 13.0	 & 19.06	 & -2.98	 & 19.29	 & \textgreater40	 & 13.2	 & 0.3	 & 14.8	 & 0.2	 & 3.1	 & 0.2	 & 2.75	 & N \\ 
E6-1-017	 & 2013-05-30	 & 15.6	 & 19.34	 & 0.12	 & 19.34	 & 24	 & 10.1	 & 0.6	 & 11.3	 & 0.4	 & 3.1	 & 0.4	 & 2.69	 & N \\ 
E6-1-037	 & 2013-05-30	 & 15.6	 & 19.36	 & -1.24	 & 19.40	 & 22	 & 8.7	 & 0.7	 & 10.2	 & 0.5	 & 2.8	 & 0.6	 & 2.65	 & N \\ 
E6-1-006	 & 2013-05-30	 & 13.9	 & 19.45	 & -1.53	 & 19.51	 & \textgreater40	 & 13.1	 & 0.4	 & 15.8	 & 0.2	 & 4.9	 & 0.2	 & 2.68	 & N \\ 
E6-1-013	 & 2013-05-30	 & 15.5	 & 19.52	 & -0.29	 & 19.52	 & 31	 & 8.3	 & 0.5	 & 9.2	 & 0.4	 & 2.7	 & 0.4	 & 2.64	 & N \\ 
E7-2-009	 & 2012-05-12	 & 15.5	 & 19.63	 & -3.26	 & 19.90	 & 29	 & 8.9	 & 0.5	 & 13.6	 & 0.3	 & 2.8	 & 0.4	 & 2.68	 & N \\ 
E7-1-024	 & 2012-05-12	 & 15.8	 & 19.87	 & -1.73	 & 19.94	 & 15	 & 9.2	 & 1.0	 & 10.1	 & 0.8	 & 0.5	 & 0.8	 & 2.66	 & N \\ 
E7-2-019	 & 2012-05-12	 & 16.4	 & 19.85	 & -3.67	 & 20.19	 & 19	 & 6.4	 & 0.9	 & 9.1	 & 0.6	 & 3.7	 & 0.6	 & 2.66	 & N \\ 
E7-2-020	 & 2012-05-12	 & 17.2	 & 19.57	 & -5.00	 & 20.20	 & 12	 & 9.6	 & 1.3	 & 12.0	 & 0.9	 & 2.4	 & 1.0	 & 2.66	 & Y \\ 
E7-1-017	 & 2012-05-12	 & 15.8	 & 20.18	 & -1.47	 & 20.23	 & 21	 & 7.0	 & 0.8	 & 14.2	 & 0.5	 & 2.2	 & 0.5	 & 2.66	 & N \\ 
E7-1-004	 & 2012-05-12	 & 12.3	 & 20.28	 & -1.08	 & 20.31	 & \textgreater40	 & 11.6	 & 0.3	 & 16.7	 & 0.2	 & 3.5	 & 0.2	 & 2.63	 & N \\ 
E7-2-008	 & 2012-05-12	 & 15.5	 & 19.70	 & -5.11	 & 20.35	 & 26	 & 9.5	 & 0.5	 & 14.2	 & 0.4	 & 4.2	 & 0.5	 & 2.66	 & N \\ 
E7-1-010	 & 2012-05-12	 & 14.7\tablenotemark{b}	 & 20.47	 & -1.19	 & 20.51	 & 39	 & 10.3	 & 0.5	 & 14.4	 & 0.3	 & 3.1	 & 0.3	 & 2.63	 & N \\ 
E7-2-036	 & 2012-05-12	 & 15.6	 & 20.32	 & -2.90	 & 20.53	 & 10	 & 12.9	 & 1.5	 & 15.5	 & 1.1	 & 3.5	 & 1.0	 & 2.68	 & N \\ 
E7-2-010	 & 2012-05-12	 & 15.8	 & 19.85	 & -5.40	 & 20.57	 & 18	 & 6.4	 & 1.0	 & 8.3	 & 0.6	 & 3.4	 & 0.6	 & 2.61	 & N \\ 
E7-1-001	 & 2012-05-12	 & 10.8	 & 20.46	 & -2.29	 & 20.59	 & \textgreater40	 & 13.8	 & 0.3	 & 20.4	 & 0.2	 & 4.2	 & 0.2	 & 2.67	 & N \\ 
E7-1-014	 & 2012-05-12	 & 15.6	 & 20.53	 & -1.66	 & 20.60	 & 19	 & 7.1	 & 0.8	 & 13.2	 & 0.6	 & 2.6	 & 0.6	 & 2.64	 & N \\ 
E7-1-008	 & 2012-05-12	 & 14.1	 & 20.47	 & -2.64	 & 20.64	 & 13	 & 14.5	 & 1.1	 & 14.8	 & 0.8	 & 4.5	 & 0.9	 & 2.69	 & N \\ 
E7-1-012	 & 2012-05-12	 & 15.3	 & 20.65	 & -0.90	 & 20.67	 & 30	 & 9.9	 & 0.4	 & 12.8	 & 0.3	 & 3.2	 & 0.4	 & 2.65	 & N \\ 
E7-2-002	 & 2012-05-12	 & 14.3	 & 20.20	 & -4.42	 & 20.68	 & \textgreater40	 & 8.6	 & 0.4	 & 12.4	 & 0.2	 & 2.9	 & 0.2	 & 2.66	 & N \\ 
E7-2-027	 & 2012-05-12	 & 18.7\tablenotemark{b}	 & 20.14	 & -4.83	 & 20.72	 & 6	 & 9.2	 & 2.6	 & 10.3	 & 1.7	 & 1.2	 & 2.0	 & 2.65	 & N \\ 
E7-2-016	 & 2012-05-12	 & 16.2	 & 19.94	 & -5.66	 & 20.73	 & 7	 & 8.0	 & 1.8	 & 12.7	 & 1.4	 & 3.1	 & 1.7	 & 2.64	 & N \\ 
E7-1-007	 & 2012-05-12	 & 13.3\tablenotemark{b}	 & 20.67	 & -2.39	 & 20.81	 & \textgreater40	 & 14.2	 & 0.3	 & 18.9	 & 0.1	 & 4.2	 & 0.2	 & 2.68	 & N \\ 
E7-2-006	 & 2012-05-12	 & 15.5	 & 20.49	 & -3.73	 & 20.82	 & \textgreater40	 & 8.4	 & 0.3	 & 10.1	 & 0.2	 & 2.6	 & 0.2	 & 2.66	 & N \\ 
E7-1-015	 & 2012-05-12	 & 15.7	 & 20.92	 & -0.95	 & 20.94	 & 22	 & 11.3	 & 0.7	 & 12.0	 & 0.5	 & 2.6	 & 0.5	 & 2.66	 & N \\ 
E7-1-022	 & 2012-05-12	 & 14.8	 & 20.98	 & -0.17	 & 20.98	 & 38	 & 8.6	 & 0.4	 & 12.7	 & 0.3	 & 3.5	 & 0.3	 & 2.67	 & N \\ 
E7-2-001	 & 2012-05-12	 & 11.5	 & 20.78	 & -2.97	 & 20.99	 & \textgreater40	 & 11.0	 & 0.3	 & 16.3	 & 0.2	 & 3.8	 & 0.2	 & 2.65	 & N \\ 
E7-1-005	 & 2012-05-12	 & 13.4	 & 20.95	 & -2.11	 & 21.06	 & 39	 & 11.9	 & 0.4	 & 14.5	 & 0.3	 & 4.0	 & 0.3	 & 2.70	 & N \\ 
E7-1-011	 & 2012-05-12	 & 15.4	 & 21.16	 & -0.76	 & 21.18	 & 33	 & 5.7	 & 0.5	 & 9.1	 & 0.3	 & 1.8	 & 0.4	 & 2.66	 & N \\ 
E7-2-023	 & 2012-05-12	 & 17.4	 & 20.75	 & -4.28	 & 21.19	 & 9	 & 6.4	 & 1.8	 & 3.2	 & 1.3	 & 0.0	 & 1.3	 & 2.69	 & N \\ 
E7-2-003	 & 2012-05-12	 & 16.0	 & 21.01	 & -3.09	 & 21.23	 & 33	 & 11.4	 & 0.5	 & 17.9	 & 0.3	 & 4.3	 & 0.4	 & 2.67	 & Y \\ 
E7-2-005	 & 2012-05-12	 & 15.0	 & 20.68	 & -5.04	 & 21.28	 & 31	 & 12.2	 & 0.4	 & 14.2	 & 0.3	 & 5.0	 & 0.3	 & 2.66	 & N \\ 
E7-1-019	 & 2012-05-12	 & 16.2	 & 21.33	 & -1.53	 & 21.38	 & 18	 & 10.6	 & 0.8	 & 11.0	 & 0.6	 & 4.0	 & 0.7	 & 2.63	 & N \\ 
E7-2-013	 & 2012-05-12	 & 16.3	 & 20.96	 & -4.36	 & 21.41	 & 29	 & 6.9	 & 0.6	 & 9.7	 & 0.4	 & 2.5	 & 0.4	 & 2.67	 & N \\ 
E7-1-003	 & 2012-05-12	 & 12.1	 & 21.40	 & -1.99	 & 21.50	 & \textgreater40	 & 15.3	 & 0.4	 & 16.9	 & 0.2	 & 5.2	 & 0.3	 & 2.70	 & N \\ 
E7-2-028	 & 2012-05-12	 & 17.3	 & 21.14	 & -4.11	 & 21.54	 & 6	 & \nodata	 & \nodata	 & \nodata	 & \nodata	 & \nodata	 & \nodata	 & 2.67	 & N \\ 
E7-2-007	 & 2012-05-12	 & 15.3	 & 21.19	 & -4.58	 & 21.68	 & 23	 & 7.3	 & 0.7	 & 9.5	 & 0.5	 & 2.6	 & 0.6	 & 2.66	 & N \\ 
E7-1-016	 & 2012-05-12	 & 15.8\tablenotemark{b}	 & 21.80	 & -0.36	 & 21.81	 & 13	 & 12.2	 & 1.2	 & 8.8	 & 1.0	 & 3.4	 & 0.8	 & 2.78	 & N \\ 
E7-1-002	 & 2012-05-12	 & 11.6	 & 21.80	 & -0.93	 & 21.82	 & \textgreater40	 & 12.4	 & 0.3	 & 15.1	 & 0.2	 & 4.0	 & 0.2	 & 2.77	 & N \\ 
E7-2-011	 & 2012-05-12	 & 15.9	 & 21.39	 & -4.61	 & 21.88	 & 31	 & 9.4	 & 0.5	 & 11.2	 & 0.3	 & 3.1	 & 0.4	 & 2.66	 & N \\ 
E7-1-021	 & 2012-05-12	 & 17.5\tablenotemark{b}	 & 21.72	 & -2.85	 & 21.91	 & 5	 & \nodata	 & \nodata	 & \nodata	 & \nodata	 & \nodata	 & \nodata	 & 2.70	 & N \\ 
E7-1-026	 & 2012-05-12	 & 16.3\tablenotemark{b}	 & 21.84	 & -1.92	 & 21.92	 & 10	 & 7.1	 & 1.5	 & 11.0	 & 1.0	 & 3.0	 & 1.0	 & 2.78	 & N \\ 
E7-2-035	 & 2012-05-12	 & 15.4	 & 21.25	 & -5.99	 & 22.08	 & 7	 & 9.3	 & 2.0	 & 6.5	 & 1.6	 & 3.8	 & 1.6	 & 2.64	 & N \\ 
E7-1-020	 & 2012-05-12	 & 16.6\tablenotemark{b}	 & 22.02	 & -1.65	 & 22.08	 & 14	 & 12.5	 & 1.0	 & 12.6	 & 0.8	 & 2.6	 & 0.9	 & 2.78	 & N \\ 
E7-2-012	 & 2012-05-12	 & 15.9	 & 21.50	 & -5.14	 & 22.10	 & 35	 & 2.9	 & 0.5	 & 5.7	 & 0.3	 & 0.9	 & 0.3	 & 2.65	 & N \\ 
E7-2-022	 & 2012-05-12	 & 16.9	 & 21.48	 & -5.54	 & 22.18	 & 9	 & 9.9	 & 1.5	 & 10.2	 & 1.3	 & 10.8	 & 1.3	 & 2.64	 & N \\ 
E7-2-015	 & 2012-05-12	 & 17.0\tablenotemark{b}	 & 21.97	 & -3.54	 & 22.26	 & 17	 & 10.9	 & 0.8	 & 12.0	 & 0.6	 & 3.5	 & 0.6	 & 2.78	 & N \\ 
E7-2-018	 & 2012-05-12	 & 17.4\tablenotemark{b}	 & 21.95	 & -3.74	 & 22.26	 & 14	 & 9.6	 & 1.0	 & 10.2	 & 0.8	 & 3.2	 & 0.8	 & 2.78	 & N \\ 
E7-1-013	 & 2012-05-12	 & 15.6\tablenotemark{b}	 & 22.32	 & -0.69	 & 22.34	 & 17	 & 11.6	 & 0.8	 & 11.7	 & 0.7	 & 4.0	 & 0.7	 & 2.78	 & N \\ 
E7-2-004	 & 2012-05-12	 & 15.4\tablenotemark{b}	 & 21.96	 & -4.12	 & 22.34	 & 34	 & 14.2	 & 0.4	 & 16.2	 & 0.3	 & 5.3	 & 0.3	 & 2.68	 & N \\ 
E7-2-017	 & 2012-05-12	 & 17.5\tablenotemark{b}	 & 21.92	 & -4.40	 & 22.35	 & 11	 & 6.9	 & 1.5	 & 9.6	 & 1.1	 & 3.2	 & 1.0	 & 2.68	 & N \\ 
E7-1-018	 & 2012-05-12	 & 16.2\tablenotemark{b}	 & 22.29	 & -2.87	 & 22.47	 & 13	 & 8.4	 & 1.0	 & 8.4	 & 0.8	 & 4.9	 & 0.8	 & 2.80	 & N \\ 
E7-1-006	 & 2012-05-12	 & 13.4\tablenotemark{b}	 & 22.73	 & -0.77	 & 22.74	 & \textgreater40	 & 11.6	 & 0.3	 & 13.5	 & 0.2	 & 4.3	 & 0.2	 & 2.78	 & N 
\enddata
\tablenotetext{a}{R.A. and decl. offset in projected distance from Sgr A* (R.A. offset is positive to the east).}
\tablenotetext{b}{Star could not be crossmatched to \citet{schodel2010}. $\emph{K}_s$ magnitude estimated from the NIFS data.}
\tablenotetext{c}{Star could not be crossmatched to \citet{schodel2010} and was too faint for reliable magnitude estimation from the NIFS data.}
\end{deluxetable*}

\begin{deluxetable*}{lcrrrrrrr}
\tabletypesize{\scriptsize}
\tablecolumns{8}
\tablecaption{NIFS Observations of Stars Without Spectral Identification}
\centering

\tablehead{    
  \colhead{Name} &
  \colhead{Date} &
  \colhead{$\emph{K}_s$} &
  \colhead{$\Delta$R.A.\tablenotemark{a}} &
  \colhead{$\Delta$Decl.\tablenotemark{a}} &
  \colhead{R ($\arcsec$)} &
  \colhead{SNR} &
  \colhead{$A_{\emph{K}_s}$} 
}
\startdata
N1-1-038	 & 2012-05-11	 & 17.3	 & 1.98	 & 7.50	 & 7.76	 & 2	 & 2.65 \\ 
N1-1-037	 & 2012-05-11	 & 17.7	 & 2.34	 & 7.63	 & 7.98	 & 3	 & 2.63 \\ 
N1-1-034	 & 2012-05-11	 & 17.1	 & 2.15	 & 7.82	 & 8.11	 & 7	 & 2.60 \\ 
N1-2-037	 & 2012-05-11	 & 17.2	 & 5.78	 & 6.42	 & 8.64	 & 6	 & 2.79 \\ 
N1-1-033	 & 2012-05-11	 & 16.2	 & 1.23	 & 8.68	 & 8.76	 & 3	 & 2.60 \\ 
N1-2-041	 & 2012-05-11	 & 17.5	 & 5.80	 & 7.61	 & 9.57	 & 4	 & 2.79 \\ 
N1-1-036	 & 2012-05-11	 & \nodata\tablenotemark{c}	 & 3.40	 & 9.36	 & 9.96	 & 0	 & 2.59 \\ 
N1-2-038	 & 2012-05-11	 & 17.1	 & 4.77	 & 8.76	 & 9.97	 & 5	 & 2.64 \\ 
N1-2-034	 & 2012-05-11	 & 16.6	 & 6.39	 & 8.11	 & 10.33	 & 2	 & 2.72 \\ 
NE-1-039	 & 2014-05-14	 & 16.5	 & 10.16	 & 2.86	 & 10.55	 & 14	 & 2.34 \\ 
N2-2-007	 & 2012-05-13	 & 15.9	 & 5.70	 & 8.93	 & 10.59	 & 0	 & 2.66 \\ 
NE-1-036	 & 2014-05-14	 & 16.3	 & 9.95	 & 4.75	 & 11.02	 & 31	 & 2.43 \\ 
NE-1-042	 & 2014-05-14	 & 17.2	 & 11.08	 & 2.70	 & 11.41	 & 6	 & 2.37 \\ 
NE-1-027	 & 2014-05-14	 & 15.9	 & 11.86	 & 2.98	 & 12.23	 & 13	 & 2.44 \\ 
N2-1-013	 & 2012-05-13	 & 15.9	 & 5.54	 & 11.34	 & 12.62	 & 3	 & 2.94 \\ 
E5-2-024	 & 2013-05-23	 & 17.3\tablenotemark{b}	 & 13.54	 & -2.87	 & 13.85	 & 5	 & 2.75 \\ 
E5-2-031	 & 2013-05-23	 & 16.9	 & 13.51	 & -3.63	 & 13.98	 & 4	 & 2.84 \\ 
E5-2-041	 & 2013-05-23	 & 18.0\tablenotemark{b}	 & 13.97	 & -1.23	 & 14.02	 & 4	 & 2.79 \\ 
E5-2-023	 & 2013-05-23	 & 15.9	 & 13.81	 & -2.60	 & 14.05	 & 15	 & 2.84 \\ 
E5-1-031	 & 2012-05-07	 & \nodata\tablenotemark{c}	 & 14.51	 & 1.20	 & 14.56	 & 0	 & 2.48 \\ 
E5-1-020	 & 2012-05-07	 & 16.6	 & 14.75	 & 1.34	 & 14.81	 & 4	 & 2.50 \\ 
E5-1-027	 & 2012-05-07	 & 17.5\tablenotemark{b}	 & 15.11	 & -0.74	 & 15.13	 & 5	 & 2.70 \\ 
E5-2-040	 & 2013-05-23	 & 16.9	 & 15.09	 & -1.49	 & 15.17	 & 13	 & 2.77 \\ 
E5-2-033	 & 2013-05-23	 & 17.7\tablenotemark{b}	 & 14.77	 & -3.46	 & 15.17	 & 6	 & 2.69 \\ 
E5-2-038	 & 2013-05-23	 & 17.5	 & 15.19	 & -2.97	 & 15.48	 & 4	 & 2.67 \\ 
E5-2-035	 & 2013-05-23	 & 18.0\tablenotemark{b}	 & 15.46	 & -2.17	 & 15.61	 & 5	 & 2.67 \\ 
E5-2-036	 & 2013-05-23	 & 17.6\tablenotemark{b}	 & 15.53	 & -2.79	 & 15.78	 & 7	 & 2.65 \\ 
E5-1-039	 & 2012-05-07	 & 18.7\tablenotemark{b}	 & 15.93	 & -1.17	 & 15.97	 & 3	 & 2.62 \\ 
E5-1-035	 & 2012-05-07	 & 18.5\tablenotemark{b}	 & 16.07	 & 0.37	 & 16.07	 & 5	 & 2.58 \\ 
E5-2-034	 & 2013-05-23	 & 17.7\tablenotemark{b}	 & 15.83	 & -2.87	 & 16.09	 & 5	 & 2.63 \\ 
E5-1-036	 & 2012-05-07	 & 17.6	 & 16.17	 & 1.48	 & 16.24	 & 6	 & 2.53 \\ 
E5-1-040	 & 2012-05-07	 & 18.7\tablenotemark{b}	 & 16.32	 & 0.15	 & 16.32	 & 5	 & 2.59 \\ 
E5-1-041	 & 2012-05-07	 & 20.1\tablenotemark{b}	 & 16.36	 & -0.23	 & 16.36	 & 1	 & 2.59 \\ 
E5-2-030	 & 2013-05-23	 & 17.2	 & 16.24	 & -2.16	 & 16.38	 & 5	 & 2.62 \\ 
E5-1-034	 & 2012-05-07	 & 17.6	 & 16.41	 & 0.84	 & 16.43	 & 14	 & 2.56 \\ 
E5-1-037	 & 2012-05-07	 & 18.6\tablenotemark{b}	 & 16.54	 & 0.43	 & 16.54	 & 4	 & 2.60 \\ 
E5-1-038	 & 2012-05-07	 & 19.1\tablenotemark{b}	 & 16.70	 & -0.02	 & 16.70	 & 1	 & 2.60 \\ 
E6-1-025	 & 2013-05-30	 & 16.5	 & 17.50	 & -0.68	 & 17.51	 & 16	 & 2.67 \\ 
E6-2-040	 & 2012-05-05	 & 17.8\tablenotemark{b}	 & 18.22	 & -3.48	 & 18.55	 & 5	 & 2.81 \\ 
E6-1-035	 & 2013-05-30	 & 18.9\tablenotemark{b}	 & 18.75	 & 0.76	 & 18.77	 & 0	 & 2.70 \\ 
E6-2-028	 & 2012-05-05	 & \nodata\tablenotemark{c}	 & 19.12	 & -2.76	 & 19.32	 & 3	 & 2.72 \\ 
E6-1-034	 & 2013-05-30	 & 18.7\tablenotemark{b}	 & 19.34	 & 0.49	 & 19.35	 & 1	 & 2.69 \\ 
E6-1-033	 & 2013-05-30	 & 17.0	 & 19.38	 & -0.86	 & 19.40	 & 6	 & 2.65 \\ 
E6-2-011	 & 2012-05-05	 & 15.6	 & 19.17	 & -3.15	 & 19.43	 & 18	 & 2.75 \\ 
E7-2-032	 & 2012-05-12	 & 17.5	 & 19.45	 & -3.37	 & 19.74	 & 3	 & 2.72 \\ 
E7-2-033	 & 2012-05-12	 & 19.5\tablenotemark{b}	 & 19.21	 & -5.14	 & 19.89	 & 3	 & 2.63 \\ 
E7-1-023	 & 2012-05-12	 & 15.6	 & 19.82	 & -1.87	 & 19.91	 & 13	 & 2.66 \\ 
E7-2-026	 & 2012-05-12	 & 17.6	 & 19.50	 & -4.26	 & 19.96	 & 4	 & 2.68 \\ 
E7-2-030	 & 2012-05-12	 & 19.5\tablenotemark{b}	 & 19.67	 & -4.20	 & 20.11	 & 2	 & 2.67 \\ 
E7-2-031	 & 2012-05-12	 & 19.4\tablenotemark{b}	 & 19.69	 & -4.64	 & 20.23	 & 4	 & 2.66 \\ 
E7-2-029	 & 2012-05-12	 & 20.6\tablenotemark{b}	 & 20.45	 & -4.77	 & 21.00	 & 1	 & 2.66 \\ 
E7-2-014	 & 2012-05-12	 & 16.3	 & 20.71	 & -4.74	 & 21.24	 & 15	 & 2.66 \\ 
E7-1-009	 & 2012-05-12	 & 14.7	 & 21.53	 & -1.91	 & 21.62	 & \textgreater40	 & 2.68 \\ 
E7-2-034	 & 2012-05-12	 & 20.4\tablenotemark{b}	 & 21.21	 & -5.11	 & 21.82	 & 1	 & 2.65 \\ 
E7-2-025	 & 2012-05-12	 & 18.7\tablenotemark{b}	 & 21.72	 & -4.00	 & 22.09	 & 5	 & 2.67 

\enddata
\tablenotetext{a}{R.A. and decl. offset in projected distance from Sgr A* (R.A. offset is positive to the east).}
\tablenotetext{b}{Star could not be crossmatched to \citet{schodel2010}. Approximate K' magnitude calculation used.}
\tablenotetext{c}{Star could not be crossmatched to \citet{schodel2010} and was too faint for magnitude approximation.}
\label{tab:unknown}
\end{deluxetable*}

\end{document}